\documentclass[fleqn,usenatbib]{mnras}

\usepackage{newtxtext,newtxmath}

\usepackage[T1]{fontenc}
\usepackage{ae,aecompl}

\usepackage{graphicx}	
\usepackage{amsmath}	

\def\simgt{\hbox{\rlap{\raise 0.425ex\hbox{$>$}}\lower 0.65ex\hbox{$\sim$}}}
\def\simlt{\hbox{\rlap{\raise 0.425ex\hbox{$<$}}\lower 0.65ex\hbox{$\sim$}}}
\def\arcsec{^{\prime\prime}}

\def\kms {km$\,$s$^{-1}$}

\def\kmsmpc {km$\,$s$^{-1}$Mpc$^{-1}$}

\def \oiii {[O{\small~III}]}

\def \nii {[N{\small~II}]}

\def \ha {H$\alpha$}

\def \re {R_{\rm e}}

\def \msun {M$_\odot$}

\def \lamr {\lambda_{\rm R}}
\def \lamre {\lambda_{\rm R_{\rm e}}}

\def \ppxf {{\small P}PXF}
\def \update {}

\def \aj {AJ}
\def \mnras {MNRAS}
\def \apj {ApJ}
\def \apjs {ApJS}
\def \apjl {ApJL}
\def \aap {A\&A}
\def \pasp {PASP}
\def \araa {ARA\&A}

\title[SAMI: disc fading]{The SAMI Galaxy Survey: The role of disc fading and progenitor bias in kinematic transitions}

\author[S. M. Croom et al.]{S.M. Croom$^{1,2,3}$\thanks{scott.croom@sydney.edu.au},
D.S. Taranu$^{3,4,5}$,
J. van de Sande$^{1,2}$,
C.D.P. Lagos$^{2,4}$,
K.E. Harborne$^{2,4}$,
\newauthor
J. Bland-Hawthorn$^{1,2}$,
S. Brough$^{6,2}$,
J.J. Bryant$^{1,2}$,
L. Cortese$^{2,4}$,
C. Foster$^{1,2}$,
M. Goodwin$^{7}$,
\newauthor
B. Groves$^{2,4,8}$,
A. Khalid$^{1}$,
J. Lawrence$^{7}$,
A.M. Medling$^{9}$,
S.N. Richards$^{10}$,
M.S. Owers$^{11,12}$,
N. Scott$^{1,2}$,
\newauthor
S.P. Vaughan$^{1,2}$
\\
$^{1}$Sydney Institute for Astronomy, School of Physics, University of Sydney, NSW 2006, Australia\\
$^{2}$ASTRO3D: ARC Centre of Excellence for All-sky Astrophysics in 3D\\
$^{3}$CAASTRO: ARC Centre of Excellence for All-sky Astrophysics\\
$^{4}$International Centre for Radio Astronomy Research, University of Western Australia, 35 Stirling Highway, Crawley, WA 6009, Australia\\
$^{5}$Department of Astrophysical Sciences, Princeton University, 4 Ivy Lane, Princeton, NJ 08544, USA\\
$^{6}$School of Physics, University of New South Wales, NSW 2052, Australia\\
$^{7}$Australian Astronomical Optics - Macquarie, 105 Delhi Rd, North Ryde, NSW 2113, Australia\\
$^{8}$Research School of Astronomy \& Astrophysics, Australian National University, Mt Stromlo Observatory, Cotter Rd, Weston Creek, ACT 2611 Australia\\
$^{9}$Ritter Astrophysical Research Center University of Toledo Toledo, OH 43606, USA\\
$^{10}$SOFIA Science Center, USRA, NASA Ames Research Center, Building N232, M/S 232-12, P.O. Box 1, Moffett Field, CA 94035-0001, USA\\
$^{11}$Department of Physics and Astronomy, Macquarie University, NSW 2109, Australia\\
$^{12}$Astronomy, Astrophysics and Astrophotonics Research Centre, Macquarie University, Sydney, NSW 2109, Australia
}

\date{Accepted XXX. Received YYY; in original form ZZZ}

\pubyear{2021}

\begin{document}
\label{firstpage}
\pagerange{\pageref{firstpage}--\pageref{lastpage}}
\maketitle

\begin{abstract}
We use comparisons between the SAMI Galaxy Survey and equilibrium galaxy models to infer the importance of disc fading in the transition of spirals into lenticular (S0) galaxies.  The local S0 population has both higher photometric concentration and lower stellar spin than spiral galaxies of comparable mass and we test whether this separation can be accounted for by passive aging alone.  We construct a suite of dynamically self--consistent galaxy models, with a bulge, disc and halo using the GalactICS code.  The dispersion-dominated bulge is given a uniformly old stellar population, while the disc is given a current star formation rate putting it on the main sequence, followed by sudden instantaneous quenching.  We then generate mock observables ($r$-band images, stellar velocity and dispersion maps) as a function of time since quenching for a range of bulge/total ($B/T$) mass ratios.  The disc fading leads to a decline in measured spin as the bulge contribution becomes more dominant, and also leads to increased concentration.  However, the quantitative changes observed after 5\,Gyr of disc fading cannot account for all of the observed difference.  We see similar results if we instead subdivide our SAMI Galaxy Survey sample by star formation (relative to the main sequence).  We use EAGLE simulations to also take into account progenitor bias, using size evolution to infer quenching time.  {\update The EAGLE simulations suggest that the progenitors of current passive galaxies typically have slightly higher spin than present day star-forming disc galaxies of the same mass.  As a result, progenitor bias moves the data further from the disc fading model scenario, implying that intrinsic dynamical evolution must be important in the transition from star-forming discs to passive discs.}
\end{abstract}




\section{Introduction}

Revealing the underlying physical processes driving the transformation of galaxies remains one of the central aims of astrophysics.  We know that through cosmic time the galaxy population tends to transition from star--forming to passive, from blue to red, and from morphologically late--type (e.g.\ spirals) to early--type (e.g.\ S0s).  These transitions are undoubtedly related to each other; for example colour is to first order related to mean stellar age, and so directly tied to the star formation history of a galaxy.  The connection between star formation history and morphology is also significant, with most star forming galaxies being late types, and most passive galaxies being early types.  However, this is not exactly a one-to-one relation, as several works have shown \citep[e.g.][]{2010MNRAS.405..783M,2009MNRAS.396..818S,2019MNRAS.483.1881D}.

Environment must play a significant role in these transformations, given the well known morphology--density \citep[e.g.][]{1980ApJ...236..351D} and star formation rate--density relations \citep[e.g.][]{2002MNRAS.334..673L}.  This is particularly so for the expected transformation from spiral to lenticular (or S0) galaxies.  The fraction of S0s grows monotonically as environment becomes richer, at the expense of spirals.  Despite environment being clearly implicated in the spiral--S0 transformation, this has not brought us directly to the physical cause of the transformation, as there remains a number of plausible mechanisms that could play a part.  In fact, it's likely that many of the proposed mechanisms have a role, but that their importance changes as a function of environment.

Measurements as a function of redshift show that as we go back in time the S0 fraction declines in dense environments.  This decline happens both in clusters \citep{1997ApJ...490..577D} and groups \citep{2010ApJ...711..192J}.  In fact, the change in S0 fraction with cosmic time appears stronger in groups (defined as having dispersion $\sigma<750$\,\kms\ by Just et al.) than clusters ($\sigma>750$\,\kms).  Similar evolution is seen in the colour \citep[e.g.][]{1984ApJ...285..426B} and star formation rates \citep[e.g.][]{2007A&A...468...33E} of galaxies in high density environments.

Arguably the simplest process that converts a spiral to an S0 is so--called strangulation \citep[e.g.][]{1980ApJ...237..692L}, where continued inflow of gas onto the disc is inhibited by the galaxy's environment.  The star formation in the disc slowly shuts down as remaining fuel is consumed.  More violent interactions, such as ram pressure stripping \citep[][]{1972ApJ...176....1G} can remove gas directly from the disc.  Ram pressure may be expected to act quickly, but as a galaxy falls into an over--dense region, the increase in ram pressure can be gradual, leading to slower transitions \citep[][]{2007MNRAS.380.1399R}.  Other physical effects {\update can also play a role.  Thermal conduction from the hot intra-cluster medium to the cooler interstellar medium of a galaxy can potentially lead to much faster gas loss \citep{2017ApJ...841...22V}.  However, simulations including magnetic fields find that thermal conduction is suppressed as the hot electrons have to follow the magnetic field lines \citep{2017ApJ...848...63V}.  Comparisons between hydrodynamic simulations of gas stripping with and without magnetic fields by \citet{2018MNRAS.476.3781R} find that gas removal is less efficient, and happens at larger radius, when magnetic fields are present.  Another contributing factor is turbulent viscosity that could enhance stripping \citep{1982MNRAS.198.1007N}, although hydrodynamical simulations seem to suggest that viscosity does not severely alter the gas mass lost from discs \citep{2008MNRAS.388L..89R}.  
}

As well as the primarily gas--physics related processes, gravitational interactions with the other galaxies or the group/cluster potential could also be important for the transition from spiral to S0.  Simulations suggest that some galaxy--galaxy mergers can lead to S0--like morphology.  These include minor mergers \citep{1998ApJ...502L.133B} and at least a fraction of major mergers with favourable impact parameters and progenitor spins \citep{2015A&A...579L...2Q}.  Less severe dynamical interactions can also play a role.   \citet[][henceforth BC11]{2011MNRAS.415.1783B} show that repeated tidal interactions with other galaxies within a group environment has the effect of heating the stellar disc, and triggering nuclear star formation to build a bulge.

Many observations of S0 galaxies have been used to try and ascertain which processes are most important.  S0 galaxies are found to follow a well defined Tully-Fisher (TF) relation \citep{2002MNRAS.330..251M,2013MNRAS.433.2667R} with an offset from the same relation for spirals.  The offset is largely consistent with S0s having older stellar populations.  However, \cite{2010MNRAS.409.1330W} finds that there remains a small offset between the spiral and S0 TF relation even when stellar mass or dynamical mass is used.  This offset may mean that galaxies undergo a small amount of contraction as they transition from S0 to spiral.  An alternative to contraction may be evolution in the zero-point of the spiral TF, although recent work carefully comparing high--redshift and low--redshift gas kinematics suggests little evolution of the TF relation \citep{2019MNRAS.482.2166T}.  The S0 TF relation therefore seems broadly consistent with gas related quenching followed by the fading of the disc, although \cite{2017A&A...604A.105T} argue that a similar TF relation could be derived through merging.

Decomposing S0 galaxies into a bulge and disc provides a different view.  \cite{2004ApJ...616..192C} suggest that S0 bulges are more luminous than can be explained by simple disc fading, but this disagrees with a combination of decomposition and colour analysis \citep{2014MNRAS.440.1690H} that is used to argue for disc fading.  Kinematic decomposition allows us to go one step further, and \cite{2013MNRAS.432.1010C} derive the TF and Faber--Jackson relations for S0 discs and bulges separately.  Their small sample shows consistent offsets of S0s in both dynamical scaling relations, that again points to more than just disc fading for the formation of S0s.  In contrast, \citet{2020MNRAS.495.4638O} have recently examined the kinematics of decomposed bulges and discs from the SAMI Galaxy Survey across a wide range in mass and morphology.  They find that the disks for both early- and late-type galaxies are sit on the same stellar-mass Tully-Fisher relation.

Measuring the stellar population ages and metallicities of Virgo cluster S0 bulges and discs separately, \citet{2014MNRAS.441..333J} find that bulges have younger ages.  This points to the last star formation in S0s being centrally concentrated, although it could still be occurring in the inner disk, rather than within a dispersion supported bulge.  The \citet{2014MNRAS.441..333J} measurement is consistent with the observation that star formation is typically more centrally concentrated in high density environments, both in clusters \citep{2004ApJ...613..866K} and groups \citep{2017MNRAS.464..121S,2019MNRAS.483.2851S}.  The younger central ages could be due to star formation enhanced by gas inflows toward the central parts of the galaxies, caused by dynamical interactions.  Alternatively, ram pressure may only remove the outer gas reservoir, allowing central star formation to continue for some time \citep{2014ApJ...781...38C}.

The advent of large-scale integral field spectroscopy surveys \citep[e.g.][]{2012MNRAS.421..872C,2015ApJ...798....7B,2016A&A...594A..36S} has opened up another window onto the question of S0 formation.  They allow estimates of the fraction of dynamical support provided by rotational
velocity ($V$) and random orbits (dispersion, $\sigma$).  These can can be combined into the spin parameter proxy, $\lambda_{R}=\langle
R|V|\rangle/\langle R\sqrt{V^2+\sigma^2}\rangle$ \citep{2011MNRAS.414..888E}, where the radius $R$ is typically taken as the effective
radius, $\re$.  \cite{2015A&A...579L...2Q} use the Calar Alto Legacy Integral Field spectroscopy Area survey (CALIFA) to argue that the transformation of spirals to S0s cannot simply be disc fading, as S0s have both lower $\lambda_R$ and higher concentration (defined as the ratio of the radii containing {\update 90 and 50 percent} of total galaxy flux, $c =R_{90}/R_{50}$).  Instead they propose that merging is able to translate galaxies in both $\lambda_R$ and concentration.  A similar conclusion is drawn using galaxies observed with the Sydney-AAO Multi-object Integral Field Spectrograph (SAMI) by \cite{2015MNRAS.454.2050F} based on cluster galaxies.  However, in this case the authors argue that the trend in $\lambda_R$ and concentration is consistent with repeated dynamical encounters (BC11).

The $\lambda_R$ vs.\ concentration plane seems to provide a useful tool for diagnosing the nature of transformations, but care has to be taken over interpretation.  Both measurements are light weighted, and so can be influenced by radial differences in stellar populations.  \cite{2016ApJ...818..180C} show that while quenched galaxies have higher Bulge/Total ($B/T$) flux ratios than star forming disc galaxies, their bulges are not more luminous.  Rather, their discs have lower luminosity.  The lower disc luminosity is a natural consequence of the disc fading as star formation ceases.  Given the bulge and disc have different light profiles (the bulge typically with higher S\'ersic index, $n$), a reduced light contribution from the disc can lead to higher measured concentration, without any underlying structural change.  Likewise, $\lambda_R$ measurements are flux weighted, so fading of a disc can lead to the bulge component dominating the measured dynamics.  If the bulge is dispersion dominated (or at least has less rotational support than the disc), then $\lambda_R$ can be reduced, again without any underlying structural change in the galaxy.

The aim of this paper is to assess how large the impact of disc fading is on $\lambda_R$ and concentration.  In particular, we wish to know whether differences between the spiral and S0 populations seen in this parameter space can be explained solely by disc fading, or if other physical effects are also required.  To do this we build self-consistent dynamical models using the GalactICS code \citep{1995MNRAS.277.1341K,2008ApJ...679.1239W}, and from them generate synthetic images and velocity fields using the MagRite code developed by \cite{2017ApJ...850...70T}.  This approach allows us to control the stellar population age of the separate dynamical components (bulge and disc).  We then compare the results of our models to integral field data from the SAMI Galaxy Survey \citep{2012MNRAS.421..872C,2015MNRAS.447.2857B}.

A challenge in comparing spirals and S0s is that we are usually making the comparison at the same redshift, while the progenitors of today's S0s were spirals at an earlier epoch.  Measurements of high redshift gas kinematics {\update appear to show} much greater turbulence in discs \citep[e.g.][]{2015ApJ...799..209W} at early times, and this could translate to higher stellar disc dispersion.
Recent simulations similarly show increased dispersion at high redshift \citep{2019MNRAS.490.3196P}.  To take this into account we will use EAGLE simulations \citep{2015MNRAS.446..521S} to make estimates of this progenitor bias.   Comparisons of star formation and kinematics using SAMI and EAGLE have already been used to highlight the importance of progenitor bias by \citet{2019MNRAS.485.2656C}.  They find that little evidence of structural change when satellite galaxies are quenched.

In Section \ref{sec:models} we describe the details of our model, including our assumed star formation histories.  In Section \ref{sec:lambdar} we present the result of making $\lambda_R$ and concentration measurements on the simulations.  Section \ref{sec:samicomp} contains a comparison of our models with measurements from the SAMI Galaxy Survey, as well as {\update discussion of the role of} progenitor bias.  We give concluding remarks in Section \ref{sec:conc}.  Throughout the paper we assume a cosmology with $\Omega_{\rm m}=0.3$, $\Omega_\Lambda=0.7$ and  $H_0=70$\,\kmsmpc.

\section{disc fading models}\label{sec:models}

Our main goal is to test whether disc fading is consistent with the difference between spirals and S0s in the $\lambda_R$--concentration plane.  To do this we need simulated galaxies that have realistic dynamics and morphological structure.  We also need to apply different star formation histories to the bulge and disc components.  Importantly, the derived kinematics need to be light weighted, so that we can fully capture the effects of only varying the $M/L$ of the stellar populations  without modifying their underlying distribution functions.  The methodology presented by \cite{2017ApJ...850...70T} to model SAMI data fulfils all of these criteria and we will now describe its key features.

\subsection{Equilibrium galaxy models}

The equilibrium galaxy models are built using a modified version of the GalactICS code \citep{1995MNRAS.277.1341K,2008ApJ...679.1239W}, detailed in Appendix F of \cite{2017ApJ...850...70T}.  GalactICS computes equilibrium phase--space distribution functions for three components: an exponential stellar disc with a sech$^2$ vertical density profile; a flattened, non-rotating \cite{1963BAAA....6...41S} profile stellar bulge; and a slightly flattened halo with a generalized \cite[][hereafter NFW]{1997ApJ...490..493N} profile.     Typically the equilibrium solution is close to the original parameters, but with the spherical components (bulge and halo) flattened by the presence of the disc.

There is a large amount of flexibility with the GalactICS approach.  However we choose a restricted range of parameters, relevant to demonstrating the impact of disc fading.  The NFW halo density profile is
\begin{equation}
\rho\propto\frac{1}{(r/r_h)^\alpha(1+r/r_h)^{(\beta-\alpha)}}, 
\end{equation}
where we choose $\alpha=1$, $\beta=2.3$ and $r_h=6.07$\,kpc. The halo is nearly spherical (mildly vertically flattened by the disc), non-rotating and truncated beyond 300\,kpc.    Modification of the halo parameters has little impact on the stellar components beyond the expected change in the rotation curve.  The bulge is also nearly spherical and non-rotating, although it can be somewhat flattened as it responds to the potential of a massive disc.  We could choose models with a rotating bulge, but a non-rotating bulge leads to the largest difference in kinematics with disc fading, so provides a robust upper limit on the role of disc fading.  The bulge follows a ``classical'' \cite{1948AnAp...11..247D} profile (S\'ersic $n_s=4.0$), although changes to the value of $n_s$ have modest impacts on our results compared to changes in the bulge scale length.  In order to generate physically realistic galaxies we use the measured stellar mass vs. $\re$ relations from \citet{2016MNRAS.462.1470L}.  They fit relations of the form
\begin{equation}
\re = a\left(\frac{M_*}{10^{10}M_\odot}\right)^b
\end{equation}
to bulge and disc properties measured from $r$-band SDSS imaging.  For bulges we use values of $a=1.667$\,kpc and $b=0.477$ to approximate the separate low- and high-mass power law relations. For discs we use $a=5.0$\,kpc and $b=0.301$ that is slightly steeper than \citet{2016MNRAS.462.1470L} to account for the difficulty of accurately measuring the size of very small discs.   The disc density profile is:
\begin{equation}
\rho\propto \exp(R/R_{d}){\rm sech}^2(z/z_d).
\end{equation}
Here $R$ is the cylindrical radius in the disc and $z$ is the vertical distance off the disc.  We choose the disc scale length, $R_{d}$, that is equivalent to an $\re$ defined using the above relations from  \citet{2016MNRAS.462.1470L}.  We assume a scale height of $z_d=0.75$\,kpc.  The structural parameter that most influences our results is the ratio of bulge and disc scale--lengths.  Changes of scale length can have important consequences for our measurements.  For example, a larger disc scale length, together with a smaller bulge would lead to large changes in $\re$ as the disc fades and the bulge becomes more important.  These changes can in turn have a significant effect on the measured $\lambda_R$.  For this reason we have chosen to use the observed relations of \citet{2016MNRAS.462.1470L} for our models.

The simulated galaxies are built by sampling the underlying distribution functions, so their spatial resolution is largely set by this discrete sampling.  The bins for sampling are adaptive.  Averaged over all bins the resolution is $\sim150$\,pc, but in practice it is better than 100\,pc in all but the outer disk.  This is an order of magnitude better than the observational resolution.

We generate a range of models with bulge/total mass fraction ($B/T$) of 0.0 to 1.0 in steps of 0.1.  In each case the total stellar mass of the combined bulge and disc is $10^{10.8}$\,\msun.

\subsection{Generating synthetic images and kinematics}

There are several steps required to simulate observed kinematics from the dynamical models presented above.  The early stages make use of the synthetic observation pipeline `This Is Not A Pipeline' \citep[TINAP; first described by][]{2013ApJ...778...61T} to generate images and kinematic maps, following the methods used to generate synthetic SAMI data described in \cite{2017ApJ...850...70T}. 

The first step is to assign a star formation history separately to the bulge and disc.  As we are primarily interested in the maximum impact that disc fading can have, we assign a uniformly old age to the bulge for all models.  The bulges are assumed to have formed in a single instantaneous burst 10\,Gyr ago.  The discs begin forming stars 12.9\,Gyr in the past with a slow exponentially declining star formation rate (SFR) $\propto\exp[-(t-t_0)/\tau]$ using a $\tau$ of 5\,Gyr.  Then we abruptly stop star formation (e.g. disc star formation instantaneously drops to zero) at times varying from 0 to 5 Gyr in the past, in 1 Gyr intervals.  We assume solar metallicity for both the bulge and disc.  Based on the SFH of each component we derive the $M/L$ ratio in three bands: SDSS $g$ and $r$, as well as an effective SAMI band over the wavelength range that we typically  measure kinematics (that we will call SAMI$gr$).  To calculate $M/L$ we use the model grids of \cite{2011MNRAS.418.2785M}.  The stellar populations are assumed to be uniform within each component (i.e.\ bulge and disc).  The dynamical masses of the bulge and disc include the contributions from evolved stars and stellar remnants (the proportion of which also varies with the SFH).

The distribution functions of the bulge and disc are then numerically integrated to create cubes of the observed luminosity density in regular spatial and projected velocity bins.  The galaxy is placed at a redshift of 0.04, giving a scale of 0.791\,kpc arcsec$^{-1}$ (typical of SAMI galaxies).  For each model we sample a range of inclinations from face--on to edge--on in steps of $15^\circ$.  

The kinematic measurements of the models are done at the native resolution of the simulations ($\sim100$\,pc, equivalent to $\sim0.13$ arcsec, an order of magnitude better than the typical observational measurements).  This is because we apply a beam-smearing correction \citep{2020MNRAS.tmp.1978H} to the  $\lamre$ measurements that we make on the SAMI data (see Section \ref{sec:sami_sample} below for details).  However, we do convolve the simulations with the expected seeing to measure concentration, as this is a seeing convolved quantity.  We also examine the impact of seeing on the kinematic measurements (see Section \ref{sec:sim_tests}).  To simulate seeing, the cube is convolved with a \cite{1969A&A.....3..455M} profile PSF, 
\begin{equation}
I(r) = I_0[1+(r/R)^2]^{-\beta_m},
\end{equation}
where the full-width at half-maximum (FWHM) is given by FWHM$=2R\sqrt{2^\frac{1}{\beta_m}-1}$.  We use FWHM=$1.8\arcsec$ {\update and $\beta_m=$2.25} [these are typical of SAMI observations and specifically based on the SAMI data of G79635, \citet{2017ApJ...850...70T}]. Similarly, the cube is convolved with a Gaussian of FWHM 150\,\kms\ along the projected velocity axis in order to match the spectral resolution of SAMI's blue arm.  The oversampled cube is then rebinned into $0.5$\,arcsec spaxels and 60\,\kms\ pixels (similar to the SAMI blue arm sampling).  The line-of-sight velocity distribution in each spaxel is fit with a Gaussian to derive the mean velocity and dispersion, with the aforementioned instrumental velocity resolution subtracted in quadrature.  It is worth noting that because we derive the luminosity density spatially and in projected velocity the derived kinematics are light weighted (to the above mentioned SAMI$gr$ band). 

The models are also projected into SDSS $g$- and $r$-band images, sampled by $0.2\arcsec$ pixels and convolved with a Moffat PSF of $1.16\arcsec$ and $0.54\arcsec$ in the $g$ and $r$ bands, respectively.  All of these seeing parameters match those used to model a SAMI galaxy (G79635) in \cite{2017ApJ...850...70T} using images from the Kilo-Degree Survey \citep[KiDS;][]{2019arXiv190211265K}. While the differential between the $g$ and $r$-band FWHMs is larger than between the median values of $0.69\arcsec$ and $0.88\arcsec$ reported in KiDS DR4, it is not unusual for weak lensing-focused optical surveys to optimize $r$-band seeing over bluer bands.

\begin{figure}
	\includegraphics[width=85mm]{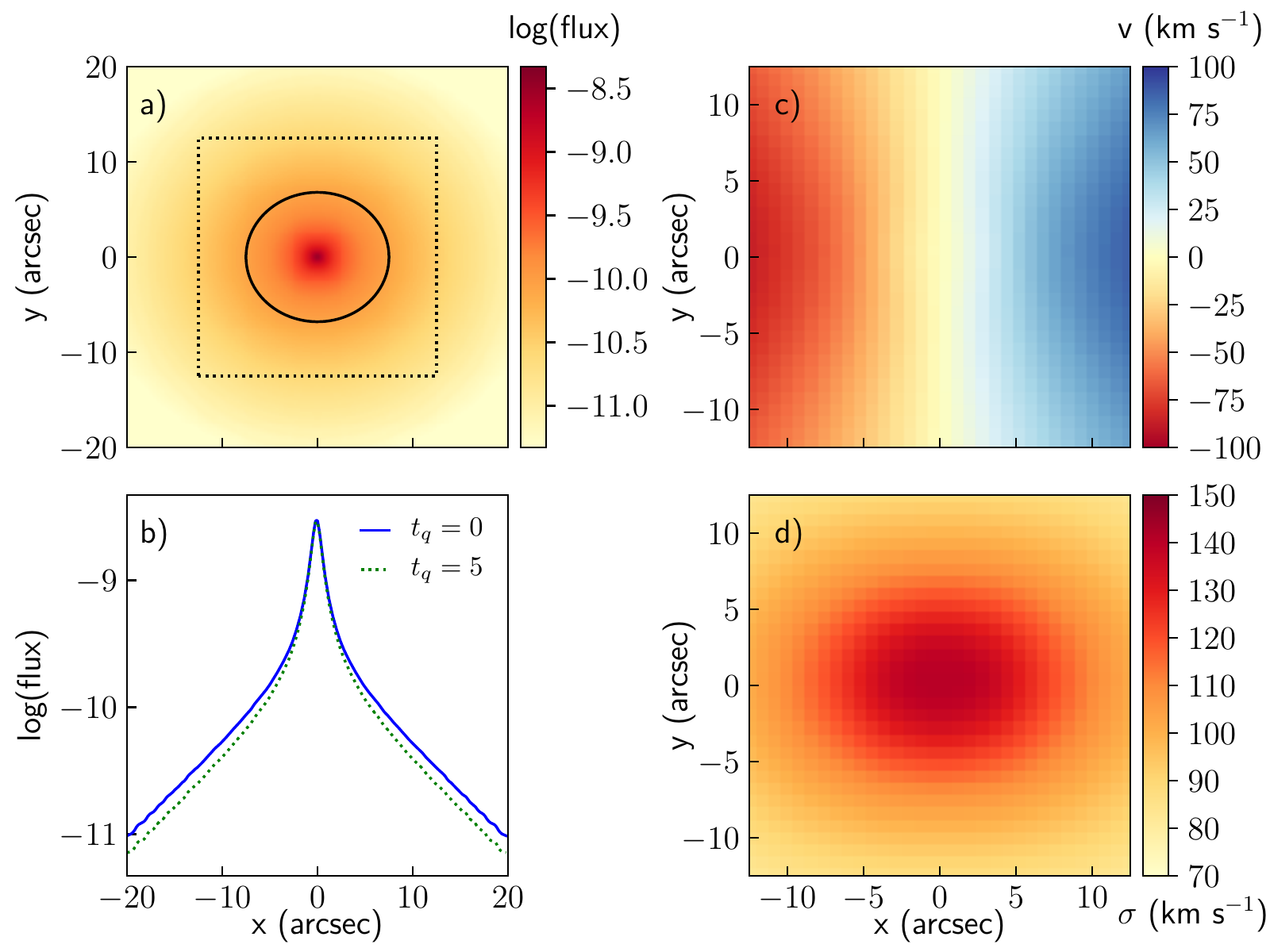}
    \caption{An example model with $B/T=0.5$, inclination = 30 degrees and $t_q=0$\,Gyr.  We show a) the $r$-band flux image on a log--scale with a 1$\re$ ellipse (solid line) and the region where the kinematics is simulated (dotted square); b) the flux profile cut along the major axis (blue line solid; units are SDSS maggies), compared to the flux profile for the same model at $t_q=5$\,Gyr (green dotted line); c) the stellar velocity map; d) the stellar velocity dispersion map.}
    \label{fig:model}
\end{figure}

An example of the simulated maps is shown in Fig.\ \ref{fig:model}.  In this case we show the $r$-band image, velocity and dispersion maps for a model galaxy with $B/T=0.5$, inclination = 30 degrees and a quenching time of $t_q=0$\,Gyr.  We also compare the flux profile to a model with $t_q=5$\,Gyr in Fig.\ \ref{fig:model}b.

\subsection{Measuring effective radius, concentration and $\lambda_R$}\label{sec:re_meas}

Our primary aim is to assess the impact of disc fading on the observed dynamical properties of galaxies.  Given this, it is important that we apply the same measurement techniques to our models as is normally applied to real observations.

\begin{figure}
	\includegraphics[width=85mm]{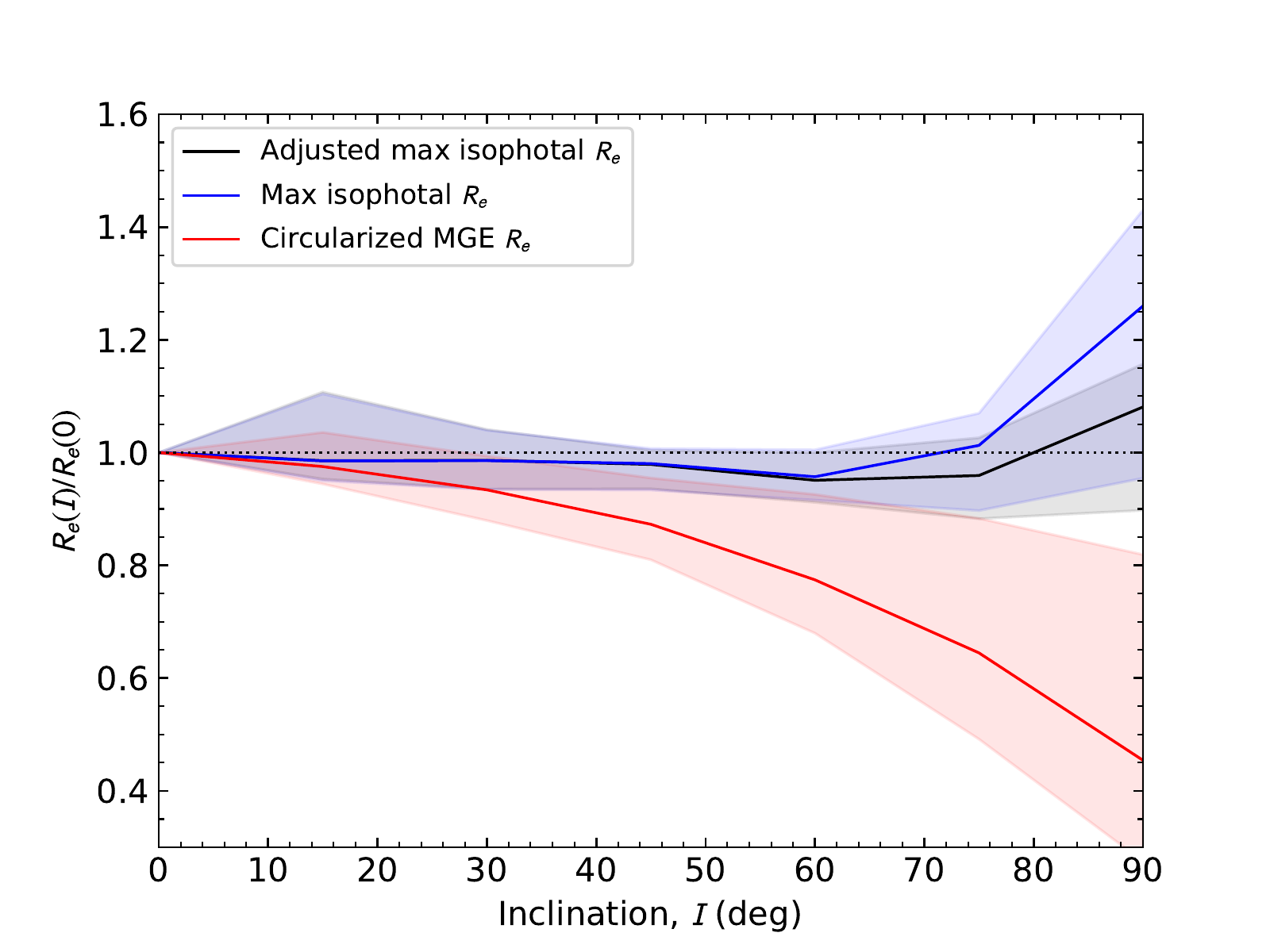}
    \caption{The variation of measured major axis $\re$ as a function of inclination, $I$.  We show the ratio of $\re$ measured from a model at a given inclination to the value for $I=0$\,deg.  The solid lines are the median over all models (full range of $B/T$ and $t_q$) as a function of inclination while the shaded regions show the full range of $\re$ ratios.  Three different $\re$ measurements are shown: a circularized MGE model (red); maximum isophotal radius (blue); adjusted maximum isophotal radius (black).}   
    \label{fig:delta_re_inc}
\end{figure}

A robust and standard approach to measuring the effective radius, $\re$, of a galaxy is via Multi-Gaussian Expansion \citep[MGE;][]{1994A&A...285..723E,2002MNRAS.333..400C}.  We apply this method to the synthetic $r$-band images, taking into account the applied seeing convolution to derive an MGE model of the unconvolved galaxy.  Our models, that are projected to a range of inclinations, allow us to test which particular $\re$ estimates are most robust.  Comparisons between different approaches are shown in Fig.\ \ref{fig:delta_re_inc}.  As pointed out by \citet{2013MNRAS.432.1709C}, circularizing the MGE model and then correcting for the ellipticity to get a major axis $\re$ value leads to systematically lower $\re$ values at high inclination (red line in Fig.\ \ref{fig:delta_re_inc}).  \citet{2013MNRAS.432.1709C} provide a much better approach that uses the MGE model of each galaxy and identifies the isophote containing half of the model light. $\re$ is then defined as the maximum radius enclosed within that isophote, analogous to the major axis of an ellipse (blue line in Fig.\ \ref{fig:delta_re_inc}).  This maximum isophotal method shows much less variation with inclination than the circularized approach, in agreement with \citet{2013MNRAS.432.1709C}.  However, at $I=90$\,deg the maximum isophotal method shows a systematic offset to give larger $\re$ values.  This offset is because in cases with a significant bulge and a thin edge--on disc the isophotes at 1$\re$ are far from elliptical.  Much less deviation is found if we adjust the maximum isophotal $\re$ so that it is the major axis radius of an ellipse with the same area and ellipticity as the half--light isophote (black line in Fig.\ \ref{fig:delta_re_inc}).  Specifically, this adjusted maximum isophotal half--light radius, $R_{\rm e,adj}$, is given by
\begin{equation}
R_{\rm e,adj} = R_{\rm e,iso} \sqrt{\frac{A_{\rm iso}}{\pi R_{\rm e,iso}^2(1-\epsilon_{\rm e,iso})}}=\sqrt{\frac{A_{\rm iso}}{\pi (1-\epsilon_{\rm e,iso})}},
\end{equation}
where $R_{\rm e,iso}$ is the maximum half--light isophotal radius, $A_{\rm iso}$ is the area within the half--light isophote, $\epsilon_{\rm e,iso}$ is the ellipticity within the half--light isophote (based on a moments of inertia analysis) and $\pi R_{\rm e,iso}^2(1-\epsilon_{\rm e,iso})$ is the area of an ellipse with semi-major axis of $R_{\rm e,iso}$ and ellipticity $\epsilon_{\rm e,iso}$.  When the half--light isophotes are close to elliptical (which is true for almost all inclinations), then $R_{\rm e,iso}$ and $R_{\rm e,adj}$ are equivalent.  Only near edge--on ($I=90$\,deg) do they diverge, with $R(I)_{\rm e,iso}$ increasingly larger than the value at $I=0$\,deg.  From here on we use $R_{\rm e,adj}$ for our half--light radius estimate for the simulations and will drop the adj subscript.  We note that using $R_{\rm e,iso}$ or $R_{\rm e,adj}$ makes no significant difference to our conclusions in this paper.

\begin{figure*}
	\includegraphics[width=1.4\columnwidth]{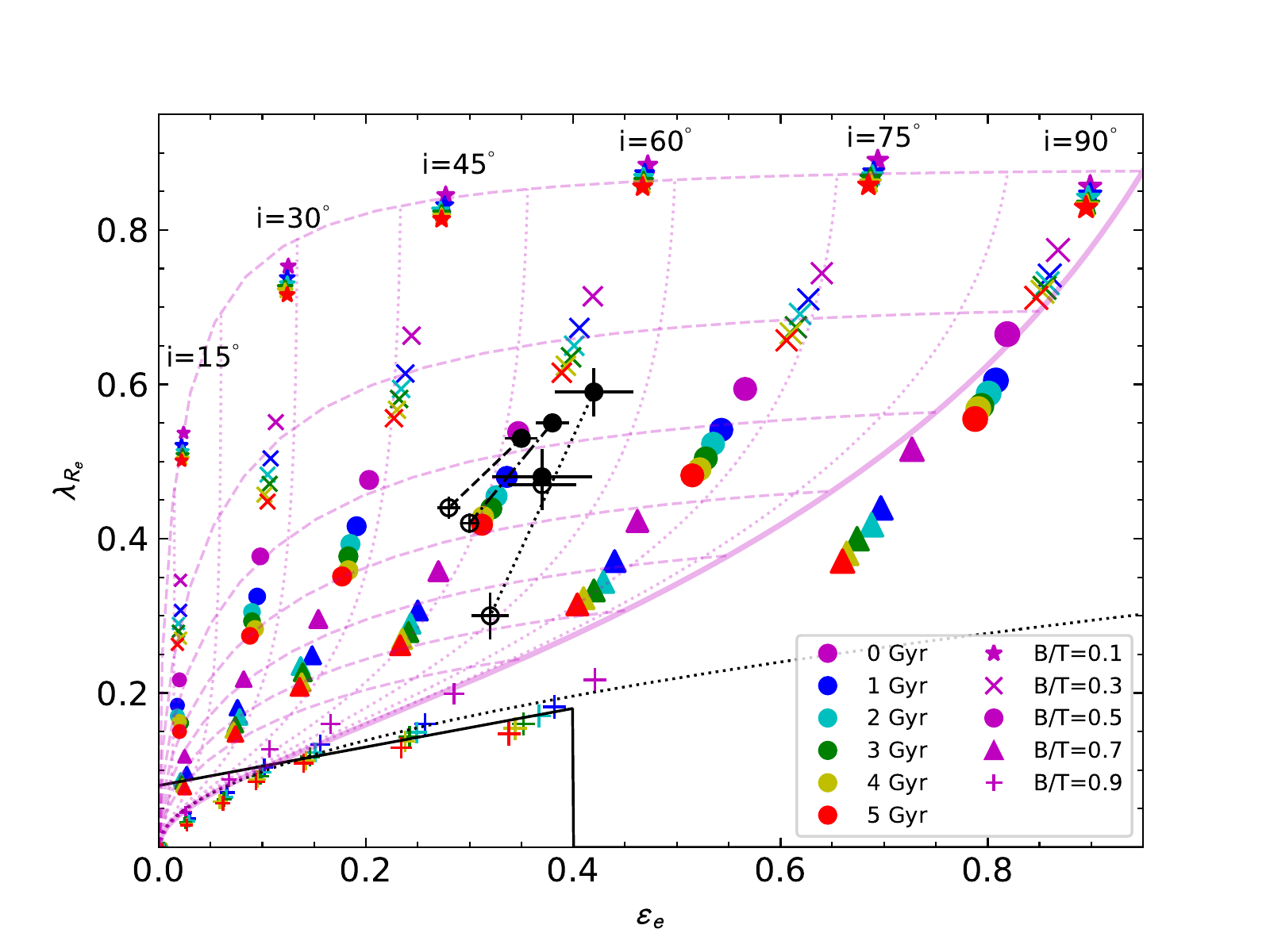}
    \caption{The distribution of our simulated galaxies in the $\lambda_R$--ellipticity plane.  We show galaxy models with $B/T=0.1$ (stars), 0.3 (crosses), 0.5 (circles), 0.7 (triangles) and 0.9 (pluses).  The colours of symbols correspond to time since quenching of 0 (magenta), 1, (blue), 2 (cyan), 3 (green), 4 (yellow), 5 (red) Gyr.  The different sets of diagonal tracks correspond to different inclinations, $i=15$, 30, 45, 60, 75, 90$^\circ$ (left to right), with point sizes becoming smaller for lower inclination.  The black lines denote typical demarcations between fast and slow rotator of $\lambda_R=0.31\sqrt(\epsilon)$ \citep[dotted line;][]{2011MNRAS.414..888E} and $\lambda_R=0.08 + 0.25\epsilon$ for $\epsilon<0.4$ \citep[solid line;][]{2016ARA&A..54..597C}.  The magenta lines show the expected relations between $\lambda_R$ and ellipticity for oblate rotators with anisotropy $\beta_z=0.7\epsilon_i$.  The solid magenta line is for edge-on galaxies with varying intrinsic ellipticity, $\epsilon_i$; the magenta dotted lines are the same model, but with different inclinations (from 10 to 80$^\circ$); the magenta dashed lines show how galaxies with fixed intrinsic ellipticity (between 0.35 and 0.85) change with inclination.  The black connected points show the mean values for early spiral, eSp, (filled circle) and S0 (empty circle) galaxies from SAMI after beam-smearing correction in the mass range $\log(M_*/M_\odot)=$9.5--10.0 (solid line), 10.0--10.5 (dashed line), 10.5--11.0 (dot-dashed line) and 11.0--11.5 (dotted line).}
    \label{fig:lam_ellip}
\end{figure*}

Various authors have compared dynamical measurements (e.g. $\lambda_R$) to galaxy concentration \citep[e.g.][]{2015MNRAS.454.2050F,2015A&A...579L...2Q}.  In these works the authors use the common SDSS definition of concentration, $C_p=r_{p,90}/r_{p,50}$, where $r_{p,50}$ and $r_{p,90}$ are the circular radii containing 50 and 90 percent of the Petrosian flux, respectively \citep{2001AJ....122.1861S}.  To maintain consistency with these previous works we also use this definition of concentration.  A concentration of $C_p\simgt2.5$ is typical of ellipticals and early--type galaxies, while $C_p\simlt2.5$ is typical for galaxies dominated by an exponential disc.  In SDSS (and other surveys) the measurement of $C_p$ is made on the seeing convolved images, so we do the same for our simulations.  The one difference is that instead of using the Petrosian flux to derive the apertures we use the total MGE model flux, and so measure $C_m=r_{m,90}/r_{m,50}$ from the seeing convolved MGE model, where $r_{m,50}$ and $r_{m,90}$ are the circular radii containing 50 and 90 percent of the total MGE model flux.  The resulting distribution of $C_m$ for the model galaxies is consistent with the observed distribution of $C_p$ from SDSS.

Finally, we measure $\lambda_R$ from the seeing convolved model kinematic maps using
\begin{equation}
\lambda_R = \frac{\sum_{i=1}^{N} F_i R_i |V_i|}{\sum_{i=1}^{N} F_i R_i\sqrt{V_i^2 + \sigma_i^2}},
\end{equation}
where the summation is over all spaxels (1 to $N$) that are contained within an elliptical aperture with semi-major axis and ellipticity of $\re$ and $\epsilon_e$ respectively.  $F_i$, $V_i$ and $\sigma_i$ are the measured flux, velocity and velocity dispersion in the $i$th spaxel.  The radius to the $i$th spaxel, $R_i$, is defined to be the semi-major axis of the ellipse that the spaxel lies on \citep[e.g.\ see][]{2016MNRAS.463..170C,2017ApJ...835..104V}.

\section{Simulation results}\label{sec:lambdar}

\subsection{Simulated $\lambda_R$ and concentration}

The suite of models described above sample inclination ($i=0$, 15, 30, 45, 60, 75, 90$^\circ$), bulge to total mass ratio (B/T=0.0 to 1.0 in steps of 0.1) and the time since quenching started ($t_q $= 0, 1, 2, 3, 4, 5\,Gyr).  In Fig.\ \ref{fig:lam_ellip} we display our models in the commonly used $\lambda_R$--ellipticity plane.  For clarity we only include models with B/T = 0.1, 0.3, 0.5, 0.7 and 0.9.  The models are compared to simple analytic descriptions of the expected tracks of oblate rotating galaxies in the $\lambda_R$--ellipticity plane that are described by  \citet{2011MNRAS.414..888E} and \citet{2007MNRAS.379..418C}, and which we will outline here for completeness.  The relation between observed ellipticity, $\epsilon$, and intrinsic ellipticity, $\epsilon_i$ is given by
\begin{equation}
\epsilon_i = 1 - \sqrt{1+\epsilon(\epsilon-2)/\sin(I)^2},
\end{equation}
where $I$ is the inclination.  \citet{2007MNRAS.379..418C} gives the approximation that
\begin{equation}
V/\sigma = \sqrt{(0.09+0.1\epsilon_i)\epsilon_i/(1-\epsilon_i)}.
\end{equation}
This can then be converted to $\lambda_R$ using the relation defined by \citet{2007MNRAS.379..401E} (see their appendix B) that incorporates a scaling factor $\kappa$, 
\begin{equation}
\lambda_R = \frac{\kappa(V/\sigma)}{\sqrt{1.0+[\kappa(V/\sigma)]^2}}.
\end{equation}
\citet{2017MNRAS.472.1272V} found that $\kappa=0.97$ for SAMI galaxies, slightly different to the value of 1.1 found by \citet{2011MNRAS.414..888E} from ATLAS3D.  This difference is based on the use of elliptical radius for SAMI measurements of $\lamre$, while ATLAS3D uses circularized radius.  The $\lambda_R$ value at a given inclination is given by
\begin{equation}
\lambda_{R,inc}  = \frac{C_i \lambda_R}{\sqrt{1+(C_i^2 -1)\lambda_R^2)}},
\end{equation}
where $C_i$ is defined by 
\begin{equation}
C_i = \frac{\sin(I)}{\sqrt{1-\beta_z \cos(I)^2}}.
\end{equation}
We take the anisotropy parameter, {\update $\beta_z=0.7\epsilon_i$ \citep{2007MNRAS.379..418C,2016ARA&A..54..597C}}.  Models tracks using the above formalism are shown in Fig.\ \ref{fig:lam_ellip} by the magenta lines.

\begin{figure}
	\includegraphics[width=1.1\columnwidth]{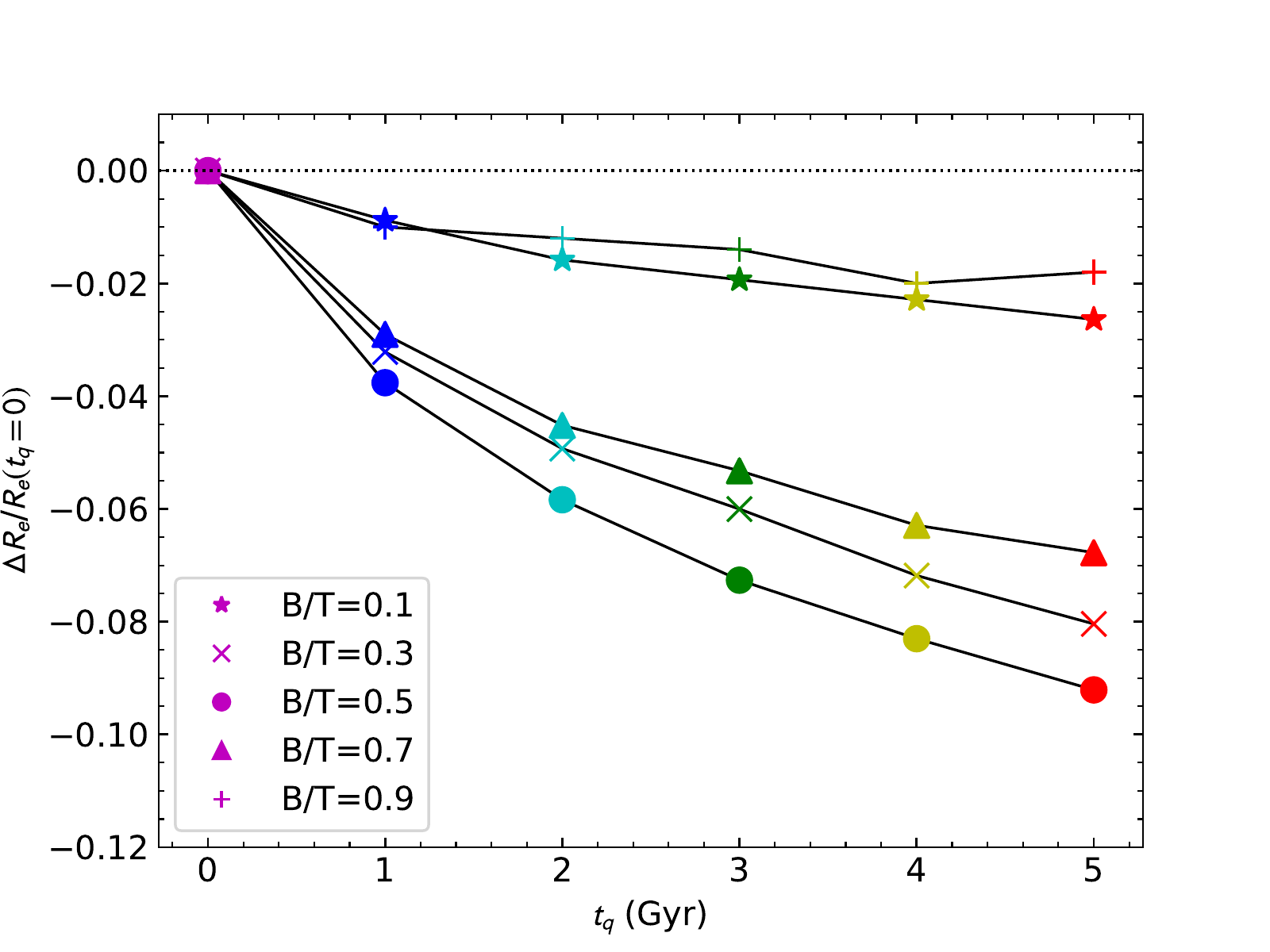}
    \caption{The fractional change in $\re$ as a function of quenching time in our models for different $B/T$ ratios.  The change in $\re$ is shown as $\Delta \re = \re(t_q)-\re(t_q=0)$, normalized by $\re(t_q=0)$.  Each point is the median over all inclinations that have been modelled. Symbol types and colours are the same as in Fig. \ref{fig:lam_ellip}.  Objects with $B/T=0.5$ show the greatest change in size.}
    \label{fig:re_tq}
\end{figure}

\begin{table}
\centering
\caption{Simulation results for the example of $B/T=0.5$ and a range of inclinations (Inc) and quenching times ($t_q$).  Other $B/T$ ratios are available in the electronic version of this paper.}\label{tab:sims}
\begin{tabular}{ccrrrrrr} 
\hline
B/T & Inc. & $t_q$  & $\re$ & $\epsilon_e$ & $\lamre$ & $(V/\sigma)_e$ & $R_{90}/R_{50}$ \\
      & deg & Gyr      & arcsec &                       &                             &                          & \\
\hline	
0.5 &  0 & 0.0 &  7.71 & 0.00 & 0.00 & 0.00 & 2.40\\
0.5 &  0 & 1.0 &  7.42 & 0.00 & 0.00 & 0.00 & 2.46\\
0.5 &  0 & 2.0 &  7.26 & 0.00 & 0.00 & 0.00 & 2.49\\
0.5 &  0 & 3.0 &  7.15 & 0.00 & 0.00 & 0.00 & 2.51\\
0.5 &  0 & 4.0 &  7.07 & 0.00 & 0.00 & 0.00 & 2.53\\
0.5 &  0 & 5.0 &  7.00 & 0.00 & 0.00 & 0.00 & 2.55\\
0.5 & 15 & 0.0 &  7.93 & 0.02 & 0.22 & 0.19 & 2.40\\
0.5 & 15 & 1.0 &  7.64 & 0.02 & 0.18 & 0.16 & 2.47\\
0.5 & 15 & 2.0 &  7.50 & 0.02 & 0.17 & 0.15 & 2.50\\
0.5 & 15 & 3.0 &  7.41 & 0.02 & 0.16 & 0.14 & 2.52\\
0.5 & 15 & 4.0 &  7.83 & 0.02 & 0.16 & 0.14 & 2.54\\
0.5 & 15 & 5.0 &  7.24 & 0.02 & 0.15 & 0.13 & 2.55\\
0.5 & 30 & 0.0 &  7.83 & 0.10 & 0.38 & 0.37 & 2.43\\
0.5 & 30 & 1.0 &  7.56 & 0.10 & 0.33 & 0.31 & 2.49\\
0.5 & 30 & 2.0 &  7.47 & 0.09 & 0.30 & 0.29 & 2.52\\
0.5 & 30 & 3.0 &  7.41 & 0.09 & 0.29 & 0.28 & 2.54\\
0.5 & 30 & 4.0 &  7.36 & 0.09 & 0.28 & 0.27 & 2.56\\
0.5 & 30 & 5.0 &  7.28 & 0.09 & 0.27 & 0.26 & 2.57\\
0.5 & 45 & 0.0 &  7.58 & 0.20 & 0.48 & 0.51 & 2.47\\
0.5 & 45 & 1.0 &  7.24 & 0.19 & 0.42 & 0.43 & 2.53\\
0.5 & 45 & 2.0 &  7.10 & 0.18 & 0.39 & 0.40 & 2.55\\
0.5 & 45 & 3.0 &  7.01 & 0.18 & 0.38 & 0.38 & 2.57\\
0.5 & 45 & 4.0 &  6.94 & 0.18 & 0.36 & 0.36 & 2.59\\
0.5 & 45 & 5.0 &  6.86 & 0.18 & 0.35 & 0.35 & 2.60\\
0.5 & 60 & 0.0 &  7.33 & 0.35 & 0.54 & 0.64 & 2.54\\
0.5 & 60 & 1.0 &  7.09 & 0.34 & 0.48 & 0.55 & 2.60\\
0.5 & 60 & 2.0 &  6.95 & 0.33 & 0.46 & 0.51 & 2.62\\
0.5 & 60 & 3.0 &  6.86 & 0.32 & 0.44 & 0.49 & 2.64\\
0.5 & 60 & 4.0 &  6.76 & 0.31 & 0.43 & 0.47 & 2.66\\
0.5 & 60 & 5.0 &  6.78 & 0.31 & 0.42 & 0.46 & 2.67\\
0.5 & 75 & 0.0 &  7.48 & 0.57 & 0.59 & 0.79 & 2.75\\
0.5 & 75 & 1.0 &  7.17 & 0.54 & 0.54 & 0.68 & 2.79\\
0.5 & 75 & 2.0 &  7.12 & 0.54 & 0.52 & 0.65 & 2.81\\
0.5 & 75 & 3.0 &  7.03 & 0.53 & 0.50 & 0.62 & 2.82\\
0.5 & 75 & 4.0 &  6.95 & 0.52 & 0.49 & 0.59 & 2.83\\
0.5 & 75 & 5.0 &  6.89 & 0.52 & 0.48 & 0.58 & 2.84\\
0.5 & 90 & 0.0 &  8.32 & 0.82 & 0.67 & 0.95 & 2.97\\
0.5 & 90 & 1.0 &  8.12 & 0.81 & 0.60 & 0.84 & 3.01\\
0.5 & 90 & 2.0 &  8.05 & 0.80 & 0.59 & 0.80 & 3.03\\
0.5 & 90 & 3.0 &  7.92 & 0.79 & 0.57 & 0.77 & 3.05\\
0.5 & 90 & 4.0 &  8.03 & 0.79 & 0.57 & 0.76 & 3.06\\
0.5 & 90 & 5.0 &  7.96 & 0.79 & 0.56 & 0.74 & 3.07\\
\hline
\end{tabular}
\end{table}

\begin{figure*}
	\includegraphics[width=1.4\columnwidth]{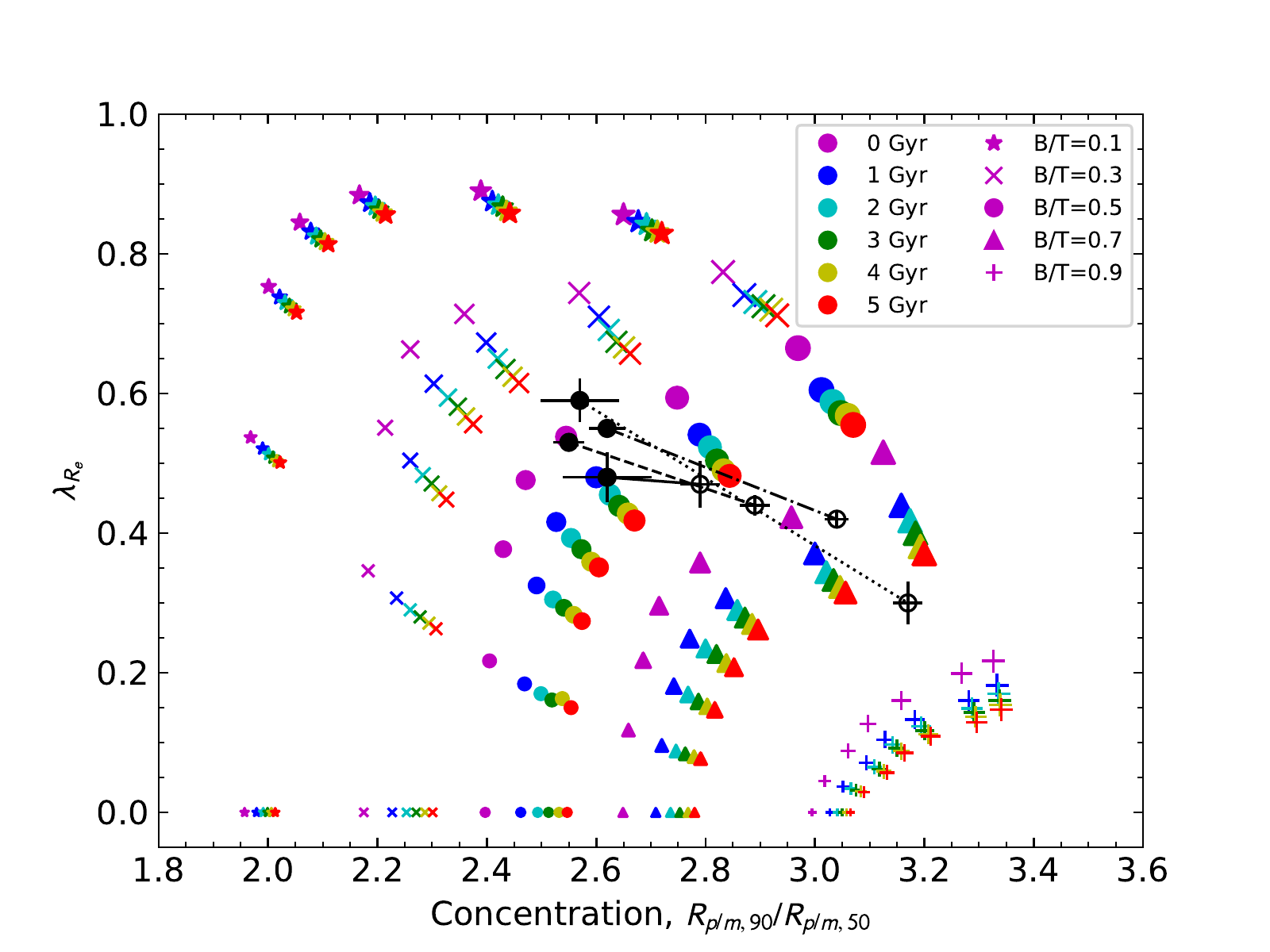}
    \caption{The distribution of our model galaxies in the $\lamre$--concentration plane.  The colours, sizes and shapes of the points are identical to Fig.\ \ref{fig:lam_ellip}.  In this figure we also include galaxies at inclination $i=0^\circ$, that all lie at $\lamre=0$.  As the inclination increases, the model tracks become steeper and have higher $\lamre$ and higher concentration.}
    \label{fig:lam_conc}
\end{figure*}

Our simulated galaxies with varying $B/T$ and quenching time are shown by the coloured points in Fig.\ \ref{fig:lam_ellip} (an example is also listed in Table \ref{tab:sims}).  Different colours denote time since quenching, $t_q$, from 0 (magenta) to 5\,Gyr (red).  Simulated galaxies with only a small bulge contribution (e.g.\ $B/T=0.1$, as indicated by a star symbol in Fig.\  \ref{fig:lam_ellip}) show little change in their location in $\lambda_R$--ellipticity  space as their disc fades.  This is as expected given that the dominant component is always the disc.  However, for galaxies with a significant bulge we see substantial variations in both their $\lambda_R$ and ellipticity as they age.  For a galaxy with $B/T=0.5$ (circles in Fig.\ \ref{fig:lam_ellip}) the change in $\lambda_R$ is $\simeq0.1$ (with some dependency on inclination) across the 5\,Gyr of disc fading that we model.  There is also a small reduction in ellipticity as the spherical bulge component contributes a larger fraction of the light at late times.  As we further increase $B/T$ the changes in $\lambda_R$ decline again, because these galaxies start with low $\lambda_R$ and ellipticity, due to being dominated by their bulge, and so the influence of the fading disc is relatively small.  The impact of disc fading is most strongly felt when the bulge and disc contribute similar amounts of light to the overall galaxy, as would be expected qualitatively.

The disc fading causes the galaxies to approximately follow the dotted magenta lines in Fig.\ \ref{fig:lam_ellip}.  These trace lines of constant inclination for a varying intrinsic ellipticity.  While these lines are derived for a single component oblate rotator, our bulge+disc models can be approximated by these lines with decreasing intrinsic ellipticity as the disc fades.    Some models fall below the limiting case of an edge--on oblate rotator (solid magenta line), and these are the edge--on cases with large dispersion--dominated bulge components. In fact, for $B/T=0.9$ our edge--on models fall into the slow rotator region, as defined by various different boundaries \citep{2011MNRAS.414..888E,2016ARA&A..54..597C}.  This is not surprising, given that our modelled bulge components are spherical and completely dispersion dominated, with no rotation.  However, we note that the edge on $B/T=0.9$ models can still have reasonably high ellipticity (up to $\epsilon_e=0.4$).  The reason for this high ellipticity is that the discs have a lower mass--to--light ratio than the bulge, and when seen edge on have higher surface brightness due to integrating through the disc.

Part of the change in measured spin in our models with time is due to a reduction in the measured $\re$ as the smaller bulge becomes more dominant as the disc fades.  The fractional quantitative change in $\re$ is shown in Fig.\ \ref{fig:re_tq}, relative to the $\re$ of each model at $t_q=0$\,Gyr.  The change in $\re$ is relatively modest, being $\simeq10$ percent after 5\,Gyr for a $B/T=0.5$.  For different $B/T$ the change in $\re$ is less than this.  We will discuss the role of size evolution further below.  

Placing our simulated galaxies in the $\lamr$--concentration plane (Fig.\ \ref{fig:lam_conc}) we see that the change in $\lambda_R$ is accompanied by a change in concentration.  For a given inclination (signified by point size), the simulated galaxies lie along a diagonal track in this plane, with the gradient depending on inclination.  Face on galaxies (small points) always have small $\lambda_R$, while edge--on galaxies show the largest change in $\lambda_R$.  The position of a galaxy along a diagonal track (for a given inclination), is mostly defined by the light--weighted B/T ratio in the SAMI$gr$ band.  This ratio is a combination of the mass--weighted B/T and the M/L of each component.  As expected given the above discussion, the largest change is for the edge--on case with a mass--weighted B/T$\simeq0.5$.  In this case the change in $\lambda_R$ is $\simeq0.1$, and the change in concentration is also $\simeq0.1$ for 5Gyr of disc fading.  As the galaxies become more face on (smaller points at lower $\lambda_R$), the change in concentration becomes larger.

We can compare our models to results from simulations that contain dynamical interactions. \citet{2015MNRAS.454.2050F}, using the simulations of BC11, find a change in $\lambda_R$ from 0.77 to 0.47 and in concentration from 2.36 to 3.71, for an edge on model that starts with a B/T ratio of 0.14 by mass.  This model  interacts with galaxies within a group for 5.6\,Gyr.  While the trends in our work and BC11 are qualitatively similar, BC11 shows larger changes in the observable parameters, particularly in concentration.  Given that the bulge mass is a small fraction of the total in the BC11 simulations, any contribution to their results from disc fading should be small.  For example, we would expect the impact to be less than the change seen in the crosses in Fig.\ \ref{fig:lam_conc}, that has a B/T of 0.3 by mass.

\subsection{Testing possible systematic effects}\label{sec:sim_tests}

The above suite of simulations provides a set of galaxies that broadly matches the observed properties of the galaxy population.  However, a number of observational and physical effects can potentially modify the measured quantities.  Here we will discuss these in turn.  Below we will generally make comparisons to our fiducial model set (described above) for a galaxy with $B/T=0.5$ and inclination of 45 degrees.

The first potential systematic we consider is the impact of atmospheric seeing.  The simulated $\lamre$ values are measured from models that have not been convolved with the seeing (as we correct our SAMI galaxy measurements for beam-smearing).  In contrast, we measure concentration after convolution with the seeing, as this matches the measurements made on the data.  To quantify the impact of these choices we compare our simulation results with and without seeing convolution.  When we do this, we find that $\lamre$ changes from 0.41 (for SAMI--like seeing) to 0.48 (native simulation resolution) for our fiducial model of $B/T=0.5$ and inclination of 45 degrees at $t_q=0$\,Gyr.  This difference is typical of the impact of seeing on $\lamre$ measurements \citep{2018MNRAS.477.4711G,2019MNRAS.483..249H,2020MNRAS.tmp.1978H} for galaxies at this spatial resolution (the ratio of seeing $\sigma$ to $\re$ is approximately $0.2$).  While there is an overall decrease in $\lamre$ in the presence of seeing, the relative evolution of $\lamre$ due to disc fading is hardly changed.  In our no seeing case 5\,Gyr of disc fading gives $\Delta\lamre=\lamre(t_q=5{\rm Gyr})-\lamre(t_q=0)=-0.116$, while for SAMI--like seeing this is $\Delta\lamre=-0.118$.  Concentration is less affected by seeing and we find that our fiducial model with $t_q=0$\,Gyr has concentration $C=2.477$ for no seeing and $C=2.471$ for SAMI-like seeing.  The change in concentration with disc fading is also unaffected with $\Delta C=C(t_q=5{\rm Gyr})-C(t_q=0)=0.135$ (for no seeing) and $\Delta C=0.134$ (for SAMI-like seeing).

The second observational effect is the $S/N$ of the measurement.   We have assumed perfect data in constructing our models, but changes in $S/N$ could impact the measured parameters.  To examine this we generate one set of galaxies (with $B/T=0.5$ and inclination of 45 degrees) with $S/N$ ratio typical of KIDS imaging and SAMI spectroscopic observations.  We then measure $\lamre$ and concentration for these simulations with added noise.  The difference caused by adding noise to the simulations is found to be at most $\simeq1.3$ percent in $\lamre$ and concentration.  We also test how the $S/N$ influences our measurement of $\Delta\lamre$ and $\Delta C$.  These are similarly small, with the difference in $\Delta\lamre$ being 0.9 percent and the difference in $\Delta C$ being 1.6 percent.  These changes are much smaller than the trends we find due to disc fading and as a result we don't consider the effect of S/N for the remainder of this paper.

Another alternative is that the disc scale height could vary with disc size.  We introduce a varying disc scale height that is one eighth of the disc scale length, varying from the fiducial value ($z_d=0.75$\,kpc) for the most massive discs to about half of this value for the least massive.  The result is only a minimal change in our results, with average changes between the fiducial model and the varying scale height model being 0.022 in $\lamre$, 0.008 in $\Delta\lamre$, $-0.045$ in concentration and $-0.011$ in $\Delta C$.

Variations in dust content between early- and late-type galaxies could influence our measurements.  The amount of extinction due to dust is found to be dependent on stellar population age \citep[e.g.][]{2008MNRAS.386.1157C}.  Disc scale-lengths have been shown to be colour dependent \citep[e.g.][]{1994A&AS..108..621P,1998MNRAS.299..595D}, and this has been explained by the distribution of dust in discs. However, more recent measurements of disc scale--lengths as a function of wavelength in large samples spanning a range of inclinations and other galaxy properties show weaker evidence for wavelength dependence \citep{2010MNRAS.406.1595F}.  Early-type galaxies can contain significant amounts of dust, but this is typically much less than late-type galaxies \citep[e.g.][]{2012ApJ...748..123S,2018MNRAS.479.1077B}.  Simulations including the impact of dust \citep[e.g.][]{2010MNRAS.403.2053G,2013A&A...557A.137P} on observed properties find that dust tends to lower the $B/T$, and make discs appear larger, with the degree of change depending on the assumed optical depth and dust geometry.  If the star-forming spirals contain significantly more dust than S0s, this would increase the observed difference in $\lamre$ and concentration.  However, given the difficulty of quantifying the differential impact of dust between spirals and S0s, we choose not to implement a dust correction in our models.

\begin{figure*}
	\includegraphics[width=2.0\columnwidth]{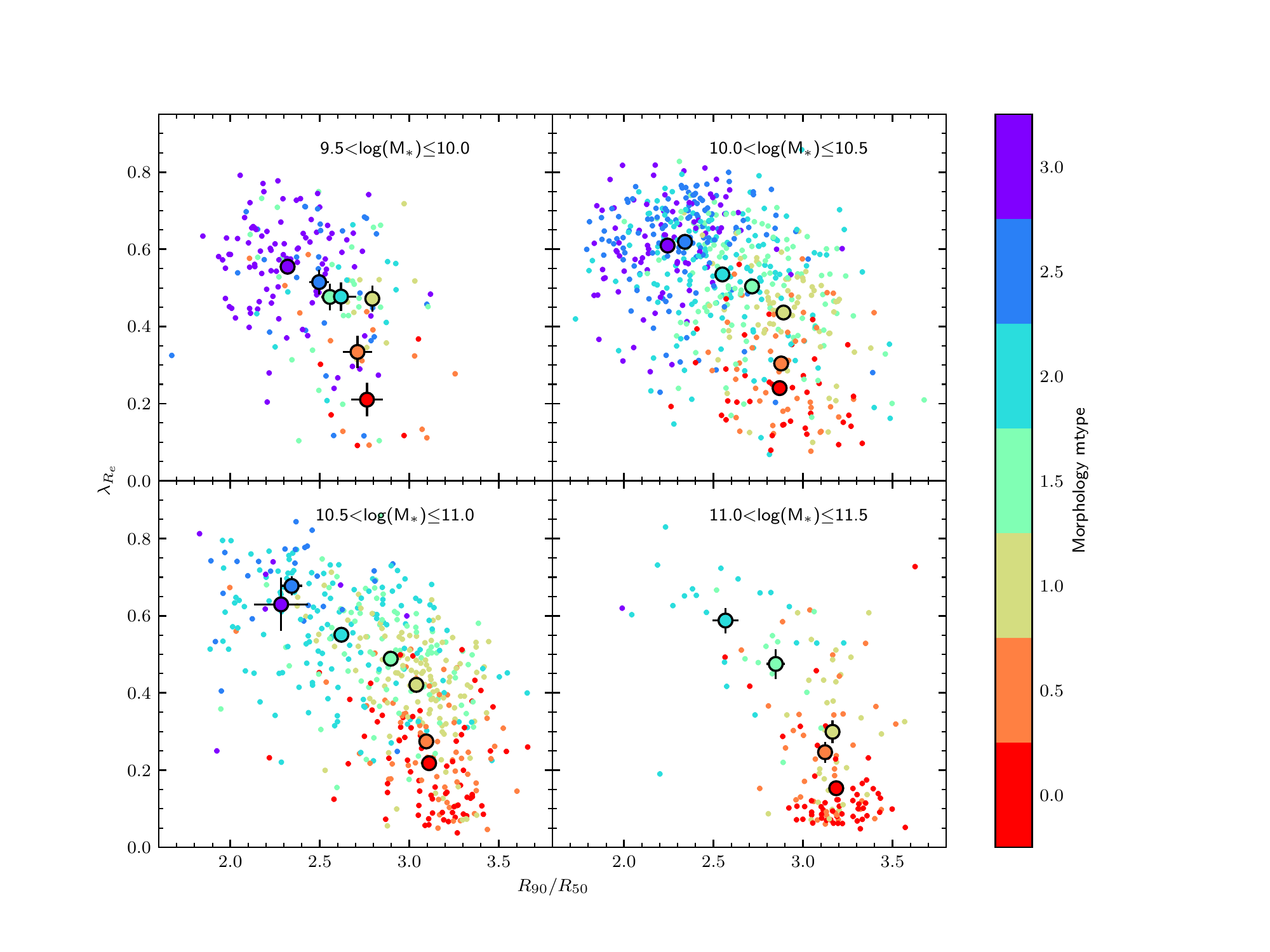}
    \caption{The distribution of our SAMI galaxies in the $\lambda_R$--concentration plane (corrected for seeing effects).  Galaxies are separated into 0.5 dex intervals of stellar mass and then colour--coded by morphology (small points) from mtype=0 (ellipticals, red points) to mtype=3 (late spirals, purple points).  The large points show the mean $\lambda_R$ and concentration values for each morphology, and are only plotted when there are at least 5 galaxies to average.  The error bars on the large points show the error on the mean and are often smaller than the points.}
    \label{fig:lam_conc_sami}
\end{figure*}

\begin{figure}
	\includegraphics[width=1.0\columnwidth]{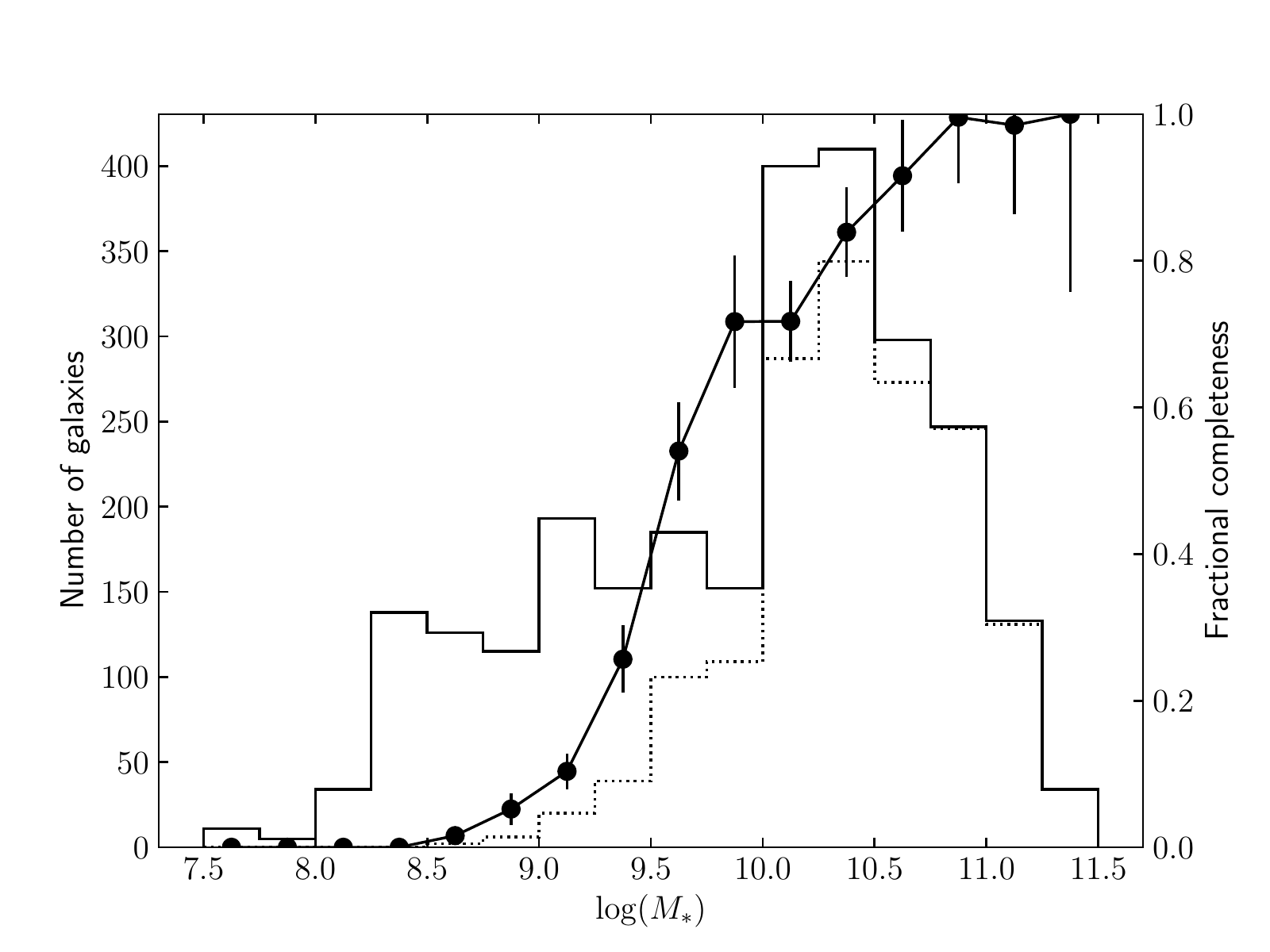}
    \caption{The stellar mass distribution of our SAMI galaxies for all objects (solid histogram) and objects for which we have robust stellar kinematics (dotted histogram).  Also shown (circles with error bars; right axis) is the fractional completeness of the stellar kinematic sample as a function of mass.}
    \label{fig:mass_comp}
\end{figure}

\section{SAMI galaxies in the $\lamre$--concentration plane}\label{sec:samicomp}

\subsection{SAMI galaxy measurements}\label{sec:sami_sample}

The Sydney--AAO Multi-object Integral field spectrograph \citep[SAMI;][]{2012MNRAS.421..872C} uses 13 deployable imaging fibre bundles \citep[hexabundles;][]{2011OpExpr.19.2649,2014MNRAS.438.869} across a 1 degree diameter field--of--view at the prime focus of the 3.9m Anglo-Australian Telescope.  The hexabundles each contains 61 fibres, and each fibre is 1.6 arcsec in diameter.  Each hexabundle therefore covers a circular 15 arcsec diameter region on the sky, with a filling factor of 75 percent.  The SAMI fibres are fed to the dual--beam AAOmega spectrograph \citep{2006SPIE.6269E..14S}.

The SAMI Galaxy Survey \citep{2012MNRAS.421..872C,2015MNRAS.447.2857B} targeted over 3000 galaxies from 2013 to 2018, covering a broad range in stellar mass ($\log(M_*/M_\odot)=10^8$ to $10^{12}$) in the redshift range $0.004<z<0.095$.  Targets were selected based on SDSS photometry and spectroscopy from the Galaxy And Mass Assembly survey \citep[GAMA;][]{2011MNRAS.413..971D}.  A further eight high density cluster regions were also targeted to capture the richest environments \citep{2017MNRAS.468.1824O}.  In the current analysis we include the SAMI cluster fields, but only those that have SDSS imaging (Abell clusters 168, 2399, 119 and 85), to maintain a consistent set of photometric measurements, particularly concentrations.

The spectroscopic observations from the SAMI Galaxy Survey cover the wavelength ranges 3750--5750\,\AA\ and 6300--7400\,\AA, at a resolution of $R=1808$ and 4304 at the spectral wavelengths of  4800\,\AA\ and 6850\,\AA, respectively \citep{2017ApJ...835..104V,2018MNRAS.481.2299S}.  

The data is reduced using a combination of the 2dFdr fibre reduction code \citep{2015ascl.soft05015A} and a purpose built pipeline \citep{2014ascl.soft07006A}.  A detailed description is given by  \cite{2015MNRAS.446.1551S} and \cite{2015MNRAS.446.1567A}.  The resulting data products are wavelength calibrated, sky subtracted and flux calibrated spectral cubes.  The cubes are generated separately for the red and blue spectrograph arms.

The simulations do not include any AGN contribution, but bright nuclear continuum from an AGN could alter the observed concentration measurements.  We visually check the nuclear spectra (3 arcsec diameter aperture) of all SAMI galaxies.  Only 11 objects from the entire sample show evidence of broad emission lines that would suggest the nucleus is not obscured.  Of these, only 2 objects (SAMI catalogue IDs 376679 and 718921) have a significant non-stellar continuum and we exclude these from our analysis below.

To compare our disc fading models to data we use stellar kinematics from the SAMI Galaxy Survey, and specifically internal data release 0.12 that contains 3071 unique galaxies from the completed survey.  These data products are identical to {\update those released in SAMI data release 3 \citep{2021MNRAS.tmp..291C}}.   Only using cluster galaxies that have SDSS imaging discounts 321 galaxies in fields with only VST-ATLAS Survey imaging \citep{2015MNRAS.451.4238S}.  As a result 597 cluster galaxies with SDSS imaging are potential objects to still include in our analysis.  

Stellar kinematics are measured using the penalized pixel fitting routine, \ppxf\  \citep{2004PASP..116..138C}, following the method discussed in detail by \cite{2017ApJ...835..104V}.  We will only highlight key points of the fitting here, and refer the reader to \cite{2017ApJ...835..104V} for further details.   The red arm data are convolved to match the blue in terms of spectral resolution and then the two arms are fitted simultaneously, assuming a Gaussian line-of-sight velocity distribution.  Optimal templates are derived by fitting in annularly binned spectra using the MILES stellar library \citep{2006MNRAS.371..703S}.  \ppxf\ is then run on individual spaxels in three passes, first to measure the noise from residuals, then to clip outlying pixels and emission lines, and finally to derive the kinematic parameters.  On the third pass \ppxf\ uses a linear combination of the optimal template in the relevant annulus and those in the adjacent annuli.  Uncertainties for each spectral measurement are estimated from fits to 150 simulated spectra, where noise is added that is consistent with the observations.

Based on the above fitting we then apply the quality cuts suggested by \cite{2017ApJ...835..104V}, namely: signal--to--noise ratio $>3\,$\AA$^{-1}$; $\sigma_{\rm obs} > {\rm FWHM}_{\rm instr}/2 \simeq 35$\,\kms; $V_{\rm error}<30$\,\kms; $\sigma_{\rm error} < \sigma_{\rm obs}*0.1 + 25$\,\kms.  The $\re$, PA and ellipticity of each SAMI galaxy is measured in the same way as the models described in Section \ref{sec:re_meas}, using MGE \citep{1994A&A...285..723E,2002MNRAS.333..400C,2009MNRAS.398.1835S}.  Detailed application of this to the SAMI data is described by \citet{2021MNRAS.tmp.1133D}.  Similarly, $\lamre$, is also measured following the procedure in Section \ref{sec:re_meas}.  We include galaxies where the $\lamre$ measurement is aperture corrected to 1$\re$ \citep{2017MNRAS.472.1272V}, in cases where the SAMI data do not extend to 1$\re$.  

For data taken at the spatial resolution of SAMI, seeing can impact the kinematic measurements.  This `beam-smearing' tends to convert velocity into dispersion and hence lower $\lamre$.  Various authors have developed beam-smearing corrections for $\lamre$ \citep[e.g.][]{2018MNRAS.477.4711G}.  We use the newly derived corrections by \citet{2020MNRAS.tmp.1978H}.  These corrections are derived by applying observational features to an array of simulated galaxies using the {\small SIMSPIN} software \citep{2020PASA...37...16H}.  The corrections are a function of $\sigma_{\rm PSF}/\re$, ellipticity and S\'ersic index where $\sigma_{\rm PSF}$ describes the width of the observational point spread function.  We only use galaxies where $\sigma_{\rm PSF}/\re<0.5$, to minimize any residual impact of beam smearing.
After correction, \citet{2020MNRAS.tmp.1978H}  find that the dispersion in $\lamre$ between the true and beam-smearing corrected simulations is only 0.026\,dex and the mean is only different by 0.001\,dex.  Beam-smearing corrections are particularly important in this work because they are dependent on galaxy size.  Early-type galaxies are on average smaller than late-type galaxies, so beam-smearing could cause systematic differences.  Applying all the kinematic quality cuts results in a sample of 1595 galaxies. 

Optical morphological classification of SAMI galaxies is described in detail by \cite{2016MNRAS.463..170C}.  The classification uses SDSS DR9 \citep{2012ApJS..203...21A} colour images inspected by at least eight independent members of the team.  First the galaxies were subdivided into early-- or late--type, based on the presence of spiral arms and/or indications of star formation.  The galaxies were then further sub-classified and given an index which we call mtype, from 0 to 3.
Early--type galaxies were further categorised as elliptical (E, mtype=0) or lenticular (S0, mtype=1) based on the presence of a disc.  Late--type galaxies were subdivided into those with a bulge (early spiral eSp, mtype=2) or without a bulge (late sprial, lSp, mtype=3).  At least 66 percent agreement was required for these classifications.  If this was not met, then adjacent votes were combined into intermediate classes with mtype = 0.5, 1.5 or 2.5.  If there is still no agreement reached, the galaxy is unclassified.  Removing galaxies that are morphologically unclassified from our kinematic sample, we then have 1566 galaxies.

Optical concentrations are taken from SDSS DR7 \citep{2009ApJS..182..543A}, and are based on the standard definition of  $C_p=r_{p,90}/r_{p,50}$, where $r_{p,50}$ and $r_{p,90}$ are the circular radii containing 50 and 90 percent of the Petrosian flux, respectively \citep{2001AJ....122.1861S}.  We find one galaxy that does not have a valid concentration (i.e. bad values or photometric flags from SDSS), resulting in a final sample of 1565 galaxies.

\subsection{Trends in $\lamre$ and concentration with morphology and mass for SAMI galaxies}

\begin{table}
\centering
\caption{The mean $\lambda_R$, concentration ($R_{90}/R_{50}$) and ellipticity ($e$) for SAMI galaxies separated by mass [in 0.5 bins of $\log(M_*/M_\odot$)] and morphological mtype.  Only bins where the number of galaxies ($N_g$) is 5 or greater are listed.}
\label{tab:sami_results}
\setlength\tabcolsep{4.0pt} 
\begin{tabular}{rrrccc} 
\hline
$\log(M_*)$ & mtype & $N_{g}$  & $\overline{\lambda_R}$ & $\overline{R_{90}/R_{50}}$ & $\overline{e}$\\
\hline
 9.5--10.0 & 0.0   &   6 & 0.210$\pm$0.044 & 2.764$\pm$0.089 & 0.100$\pm$0.026\\
 9.5--10.0 & 0.5   &  15 & 0.334$\pm$0.042 & 2.711$\pm$0.083 & 0.127$\pm$0.023\\
 9.5--10.0 & 1.0   &  11 & 0.472$\pm$0.034 & 2.793$\pm$0.038 & 0.369$\pm$0.035\\
 9.5--10.0 & 1.5   &  27 & 0.476$\pm$0.034 & 2.556$\pm$0.047 & 0.298$\pm$0.032\\
 9.5--10.0 & 2.0   &  11 & 0.477$\pm$0.037 & 2.618$\pm$0.084 & 0.366$\pm$0.051\\
 9.5--10.0 & 2.5   &  29 & 0.515$\pm$0.032 & 2.496$\pm$0.054 & 0.393$\pm$0.037\\
 9.5--10.0 & 3.0   &  96 & 0.555$\pm$0.013 & 2.320$\pm$0.025 & 0.417$\pm$0.020\\
10.0--10.5 & 0.0   &  38 & 0.240$\pm$0.019 & 2.870$\pm$0.045 & 0.074$\pm$0.010\\
10.0--10.5 & 0.5   &  48 & 0.304$\pm$0.018 & 2.878$\pm$0.030 & 0.131$\pm$0.012\\
10.0--10.5 & 1.0   &  66 & 0.436$\pm$0.015 & 2.891$\pm$0.028 & 0.283$\pm$0.011\\
10.0--10.5 & 1.5   & 112 & 0.504$\pm$0.013 & 2.715$\pm$0.029 & 0.345$\pm$0.014\\
10.0--10.5 & 2.0   & 148 & 0.535$\pm$0.012 & 2.550$\pm$0.028 & 0.351$\pm$0.016\\
10.0--10.5 & 2.5   & 126 & 0.619$\pm$0.011 & 2.340$\pm$0.024 & 0.436$\pm$0.021\\
10.0--10.5 & 3.0   &  85 & 0.610$\pm$0.013 & 2.244$\pm$0.029 & 0.417$\pm$0.025\\
10.5--11.0 & 0.0   &  74 & 0.218$\pm$0.014 & 3.111$\pm$0.029 & 0.103$\pm$0.008\\
10.5--11.0 & 0.5   &  64 & 0.274$\pm$0.017 & 3.095$\pm$0.035 & 0.144$\pm$0.009\\
10.5--11.0 & 1.0   & 118 & 0.421$\pm$0.013 & 3.040$\pm$0.022 & 0.299$\pm$0.009\\
10.5--11.0 & 1.5   &  90 & 0.489$\pm$0.014 & 2.896$\pm$0.029 & 0.380$\pm$0.016\\
10.5--11.0 & 2.0   & 138 & 0.551$\pm$0.012 & 2.621$\pm$0.033 & 0.379$\pm$0.016\\
10.5--11.0 & 2.5   &  27 & 0.677$\pm$0.025 & 2.343$\pm$0.058 & 0.478$\pm$0.040\\
10.5--11.0 & 3.0   &   7 & 0.630$\pm$0.069 & 2.284$\pm$0.151 & 0.291$\pm$0.095\\
11.0--11.5 & 0.0   &  59 & 0.153$\pm$0.016 & 3.186$\pm$0.025 & 0.112$\pm$0.009\\
11.0--11.5 & 0.5   &  36 & 0.246$\pm$0.027 & 3.123$\pm$0.032 & 0.224$\pm$0.017\\
11.0--11.5 & 1.0   &  31 & 0.299$\pm$0.030 & 3.165$\pm$0.027 & 0.319$\pm$0.018\\
11.0--11.5 & 1.5   &  11 & 0.475$\pm$0.039 & 2.848$\pm$0.052 & 0.323$\pm$0.043\\
11.0--11.5 & 2.0   &  20 & 0.588$\pm$0.033 & 2.567$\pm$0.072 & 0.416$\pm$0.038\\
\hline
\end{tabular}
\end{table}

The distribution of SAMI galaxies in the $\lamre$--concentration plane is shown in Fig. \ref{fig:lam_conc_sami}.  To distinguish mass trends from other effects we separate the galaxies into bins of 0.5 dex in $\log(M_*/M_\odot)$.  At masses below $\log(M_*/M_\odot)=10$ there are a smaller number of galaxies and the range of morphologies is more limited.  One aspect of this is the purely physical effect that most low-mass galaxies are late-type spirals or irregulars.  However, another factor is that the lower surface brightness for these galaxies means that the 
measured stellar kinematics is less complete at low masses.  In Fig. \ref{fig:mass_comp} we show the stellar kinematic completeness as a function of mass and below $\log(M_*/M_\odot)=9.5$ this drops quickly.  However, above $\log(M_*/M_\odot)=10.0$ our stellar kinematic measurements are relatively complete, and we also have a broad range of morphology.  The above points highlight the need to make consistent comparisons at the same stellar mass when investigating morphology trends.  

\begin{table*}
\centering
\caption{The difference between mean $\lamre$ and concentration ($C$) for different morphological or SFR-defined classes.  This is shown in mass intervals for the difference between S0 (mtype=1) and eSp (mtype=2) galaxies, the different between INT and SF galaxies, and the difference between PAS and SF galaxies.  Samples listed as 'No-SR' do not include slow rotators [that lie within the region defined by \citet{2016ARA&A..54..597C}].   We also list the significance of the difference and the fractional difference that can be contributed by disc fading.  For the disc fading contribution we compare to the model results averaged over all B/T and inclination values, of $\Delta \lamre=-0.056$ and $\Delta C=0.091$.  The last row for each sample contains the results for the full mass range between $\log(M_*/M_\odot)=9.5$ and 11.5, based on an inverse variance weighted average of the individual mass bins.}
\label{tab:sami_delta}
\setlength\tabcolsep{4.0pt} 
\begin{tabular}{rcccccc} 
\hline
& $\Delta\lamre$ & $\Delta C$ & \multicolumn{2}{c}{Significance} & \multicolumn{2}{c}{Frac. DF}\\
$\log(M_*)$ & (S0-eSp) &(S0-eSp) & $\Delta\lamre$ & $\Delta$C & $\Delta\lamre$ & $\Delta C$ \\
\hline	
\multicolumn{7}{c}{All Morph (S0-eSp)}\\
 9.5--10.0  &  $-$0.006$\pm$0.048  &   0.175$\pm$0.088 & $-$0.1 &  2.0 &  9.93 &  0.47\\ 
10.0--10.5  &  $-$0.099$\pm$0.020  &   0.341$\pm$0.039 & $-$5.0 &  8.7 &  0.56 &  0.24\\ 
10.5--11.0  &  $-$0.130$\pm$0.018  &   0.419$\pm$0.040 & $-$7.3 & 10.5 &  0.42 &  0.20\\ 
11.0--11.5  &  $-$0.288$\pm$0.044  &   0.598$\pm$0.075 & $-$6.6 &  7.9 &  0.19 &  0.14\\ 
All  &  $-$0.132$\pm$0.013  &   0.417$\pm$0.026 & $-$10.4 & 16.1 &  0.42 &  0.20\\ 
\hline
\multicolumn{7}{c}{No-SR Morph (S0-eSp)}\\
 9.5--10.0  &  $-$0.006$\pm$0.048  &   0.175$\pm$0.088 & $-$0.1 &  2.0 &  9.93 &  0.47\\ 
10.0--10.5  &  $-$0.097$\pm$0.017  &   0.352$\pm$0.039 & $-$5.7 &  9.0 &  0.56 &  0.23\\ 
10.5--11.0  &  $-$0.098$\pm$0.016  &   0.409$\pm$0.040 & $-$6.2 & 10.1 &  0.56 &  0.20\\ 
11.0--11.5  &  $-$0.220$\pm$0.037  &   0.596$\pm$0.079 & $-$5.9 &  7.6 &  0.25 &  0.14\\ 
All  &  $-$0.104$\pm$0.011  &   0.409$\pm$0.026 & $-$9.6 & 15.5 &  0.53 &  0.20\\ 
\hline
\multicolumn{7}{c}{No-SR SFR (INT-SF)}\\
 9.5--10.0  & $-$0.026$\pm$0.033 &  0.100$\pm$0.054 &  $-$0.8 &   1.8  &   2.14 &   0.91\\ 
10.0--10.5  & $-$0.066$\pm$0.018 &  0.194$\pm$0.036 &  $-$3.6 &   5.4  &   0.85 &   0.47\\ 
10.5--11.0  & $-$0.084$\pm$0.019 &  0.301$\pm$0.045 &  $-$4.3 &   6.6  &   0.67 &   0.30\\ 
11.0--11.5  & $-$0.210$\pm$0.038 &  0.403$\pm$0.104 &  $-$5.5 &   3.9  &   0.27 &   0.23\\ 
All         & $-$0.081$\pm$0.011 &  0.309$\pm$0.026 &  $-$7.3 &  12.1  &   0.69 &   0.29\\ 
\hline
\multicolumn{7}{c}{No-SR SFR (PAS-SF)}\\
 9.5--10.0 & $-$0.105$\pm$0.022 &  0.272$\pm$0.042 &  $-$4.7 &   6.4  &   0.53 &   0.34\\ 
10.0--10.5 & $-$0.194$\pm$0.012 &  0.445$\pm$0.026 & $-$15.9 &  16.8  &   0.29 &   0.20\\ 
10.5--11.0 & $-$0.191$\pm$0.016 &  0.517$\pm$0.035 & $-$12.3 &  14.8  &   0.29 &   0.18\\ 
11.0--11.5 & $-$0.307$\pm$0.037 &  0.603$\pm$0.098 &  $-$8.3 &   6.1  &   0.18 &   0.15\\ 
All        & $-$0.187$\pm$0.008 &  0.506$\pm$0.019 & $-$22.6 &  26.8  &   0.30 &   0.18\\ 
\hline
\end{tabular}
\end{table*}

When we subdivide our sample by morphology (mtype) at $\log(M_*/M_\odot)>9.5$ (colour coded in Fig. \ref{fig:lam_conc_sami} from purple to red for late--type to early--type) we see the expected trends that earlier galaxy types (lower mtype value) have higher concentration and lower $\lamre$.  While there is significant scatter (in part caused by inclination), the mean trends (large points) are clear.  The mean values are also given in Table \ref{tab:sami_results} (note that we only give mean values when we have at least 5 galaxies in a mass--morphology bin).

Our main aim in this paper is to test whether the changes seen when spirals transition to S0 galaxies could be consistent with disc fading.  For this it is best to define samples that are minimally contaminated with other morphologies, that could bias our measurements.  We therefore now consider only those objects for which the morphology is classified as S0 with mtype=1 (yellow points in Fig.\ \ref{fig:lam_conc_sami}).  We do not include objects with morphological classifications of mtype=0.5 or 1.5, as these intermediate classes were only assigned when agreement could not be reached in our classification.  Therefore, they likely contain contamination from adjacent morphological classes.

As the comparison to our S0s we take the SAMI objects classified as pure early-type spiral galaxies, mtype=2 (light blue points in Fig.\ \ref{fig:lam_conc_sami}).  We make this choice because this class should be minimally contaminated by S0s, have significant numbers across each of the mass intervals above $\log(M_*/M_\odot)=10.0$, and based on their classification should show evidence for a bulge.  That said, we note that the mtype=2.5 or 3.0 classes are generally close to the mtype=2.0 galaxies in the $\lamre$--concentration plane.  If anything, the later mtypes have slightly higher $\lamre$ and lower concentration, so looking at the difference between mtype=2.0 (early spiral, eSp) and mtype=1.0 (S0) provides a lower limit on the global difference between spirals and S0s.

The values of  $\Delta\lamre = \lamre({\rm eSp}) - \lamre({\rm S0})$ and $\Delta C = C(eSp)-C(S0)$ are listed in Table \ref{tab:sami_delta}.  In all three mass intervals above $\log(M_*/M_\odot)=10.0$ the difference in concentration and $\lamre$ between eSp (mtype=2) and S0 (mtype=1) galaxies is highly significant.   Taking the average across the full mass range ($\log(M_*)=9.5$ to 11.5, although limiting to greater than 10.0 make no difference to the results) gives a mean $\Delta\lamre=-0.132\pm0.013$ and mean $\Delta C=0.417\pm0.026$.  In calculating the average across all masses we take an inverse variance weighted average of the differences in each of the four separate mass intervals.  This approach is more reliable than calculating the mean $\lamre$ and concentration using a single large mass bin, as the changing morphological mix as a function of mass can bias the difference in this case. 

There is a trend of increasing difference between eSp and S0 galaxies as mass increases for both $\lamre$ and concentration.  In $\lamre$ the trend with mass is largely driven by decreasing spin for S0 galaxies as mass increases.  However, the main trend in concentration is increasing concentration as mass increases for S0 galaxies (Table \ref{tab:sami_delta}).  A natural interpretation of increasing concentration with mass is that higher mass S0s have a larger B/T ratio.  We do indeed see this trend in bulge-disc decomposition of GAMA galaxies (Casura et al. in prep).  However, the fact that we don't see equivalent changes in $\lamre$ suggests that the bulges (or at least more concentrated components) still have substantial dynamical support from rotation.  Future work will focus on explicit kinematic bulge-disc decomposition to explore this issue further.

\begin{figure}
	\includegraphics[width=1.2\columnwidth]{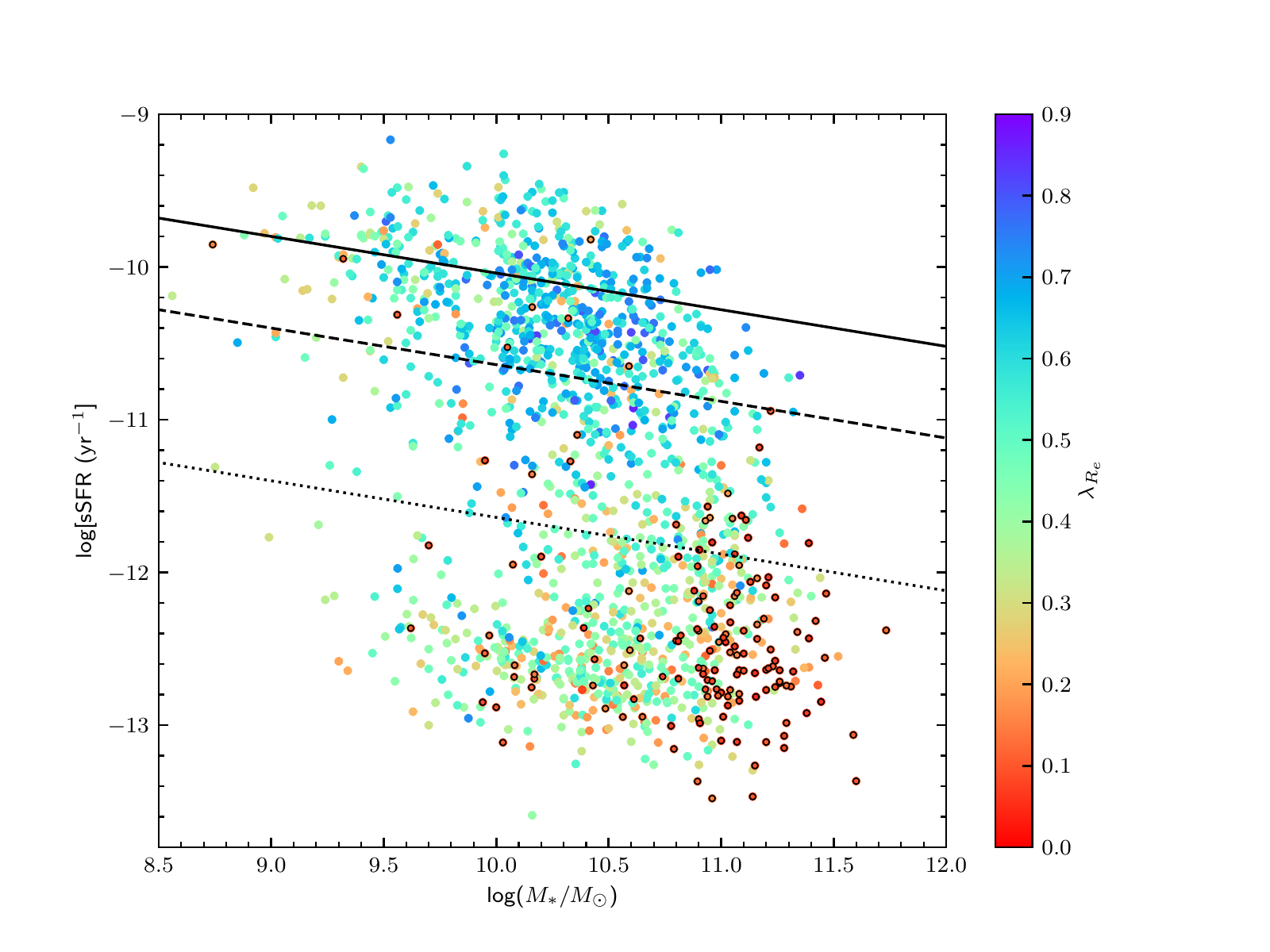}
    \caption{The distribution of SAMI galaxies with $\lamre$ measurements in the sSFR-mass plane.  Points are colour-coded by their $\lamre$ value.  Points with black edges are classified as slow rotators based the \citet{2016ARA&A..54..597C} definition.  Diagonal lines indicate the location of the star formation main sequence from \citet{2015ApJ...801L..29R} (solid line); 0.6 dex below the main sequence (dashed line); 1.6 dex below the main sequence (dotted line).}
    \label{fig:sfr_ms}
\end{figure}

Another possible contribution to the mass trends seen is the increased fraction of slow rotators above $\log(M_*/M_\odot)=11.0$ \citep[e.g.][]{2017ApJ...844...59B,2017MNRAS.472.1272V}.  The S0 sample could be contaminated by slow rotators at high mass, leading to lower spin measurement.  At lower masses this contamination is unlikely to be causing any offsets, given that the fraction of slow rotators is small.  To directly test this, we recalculate our SAMI results but removing any galaxy that is within the slow rotator region defined by $\lamre<0.08+\epsilon/4$ and $\epsilon<0.4$ \citep{2016ARA&A..54..597C}.  The results of removing the slow rotators are shown in Table \ref{tab:sami_delta}.  Unsurprisingly, the eSp galaxies are unaffected by the slow rotator cut, as only 3 out of 317 eSp galaxies lie in the slow rotator region of the $\lamre$--ellipticity plane (face-on disks).  The S0 galaxies are somewhat more impacted by removing slow rotators, particularly at high mass.  This is either due to true S0s with lower average $\lamre$ scattering into the slow rotator region, or intrinsically slow-rotating galaxies being misclassified as S0s as part of our morphological classification.  Removing the slow rotators leads to a smaller overall change in $\lamre$ with $\Delta\lamre=-0.104\pm0.011$ (slow rotators removed) compared to $-0.129\pm0.013$ (retaining slow rotators) when averaged over all masses.  In contrast the concentrations are hardly affected with $\Delta C=0.409\pm0.026$ (slow rotators removed) compared to $0.417\pm0.026$ (retaining slow rotators).  For both $\lamre$ and concentration the difference between eSp and S0 galaxies is still highly significant, even when removing slow rotators.

\begin{table*}
\centering
\caption{The mean $\lambda_R$, concentration ($R_{90}/R_{50}$) and ellipticity ($e$) for SAMI galaxies separated by mass [in 0.5 bins of $\log(M_*/M_\odot$)] and star formation rate.  Galaxies are separated into star formation rate classes (SF, intermediate, passive) based on their location with respect to the star-forming main sequence.  The mean specific star formation rate for each class is also listed, although we note that values below $\log(sSFR)\simeq-12.0$ are not reliable and should be regarded as non-detections.  Only bins where the number of galaxies ($N_g$) is 5 or greater are given.}
\label{tab:sami_results_sfr}
\setlength\tabcolsep{4.0pt} 
\begin{tabular}{rccrccc} 
\hline
$\log(M_*/M_\odot)$ & SF class & $\overline{\log(sSFR)}$ & $N_{g}$  & $\overline{\lambda_R}$ & $\overline{R_{90}/R_{50}}$ & $\overline{e}$\\
& & (yr$^{-1}$) &&\\
\hline
 9.5--10.0 & SF    & $-$10.06 & 118 & 0.542$\pm$0.013 & 2.372$\pm$0.026 & 0.408$\pm$0.018\\
 9.5--10.0 & INT   & $-$11.04 &  27 & 0.516$\pm$0.031 & 2.472$\pm$0.049 & 0.303$\pm$0.036\\
 9.5--10.0 & PAS   & $-$12.45 &  54 & 0.437$\pm$0.018 & 2.644$\pm$0.034 & 0.282$\pm$0.023\\
10.0--10.5 & SF    & $-$10.21 & 328 & 0.591$\pm$0.007 & 2.396$\pm$0.017 & 0.399$\pm$0.012\\
10.0--10.5 & INT   & $-$11.21 &  92 & 0.525$\pm$0.017 & 2.590$\pm$0.032 & 0.344$\pm$0.020\\
10.0--10.5 & PAS   & $-$12.53 & 189 & 0.397$\pm$0.010 & 2.841$\pm$0.020 & 0.228$\pm$0.011\\
10.5--11.0 & SF    & $-$10.44 & 134 & 0.580$\pm$0.013 & 2.526$\pm$0.030 & 0.371$\pm$0.017\\
10.5--11.0 & INT   & $-$11.46 & 121 & 0.496$\pm$0.014 & 2.827$\pm$0.034 & 0.334$\pm$0.018\\
10.5--11.0 & PAS   & $-$12.50 & 211 & 0.389$\pm$0.009 & 3.044$\pm$0.018 & 0.259$\pm$0.010\\
11.0--11.5 & SF    & $-$10.75 &  12 & 0.644$\pm$0.031 & 2.544$\pm$0.096 & 0.376$\pm$0.049\\
11.0--11.5 & INT   & $-$11.63 &  41 & 0.434$\pm$0.024 & 2.946$\pm$0.049 & 0.281$\pm$0.028\\
11.0--11.5 & PAS   & $-$12.45 &  37 & 0.337$\pm$0.022 & 3.147$\pm$0.036 & 0.264$\pm$0.022\\
\hline
\end{tabular}
\end{table*}

\begin{figure*}
	\includegraphics[width=2.0\columnwidth]{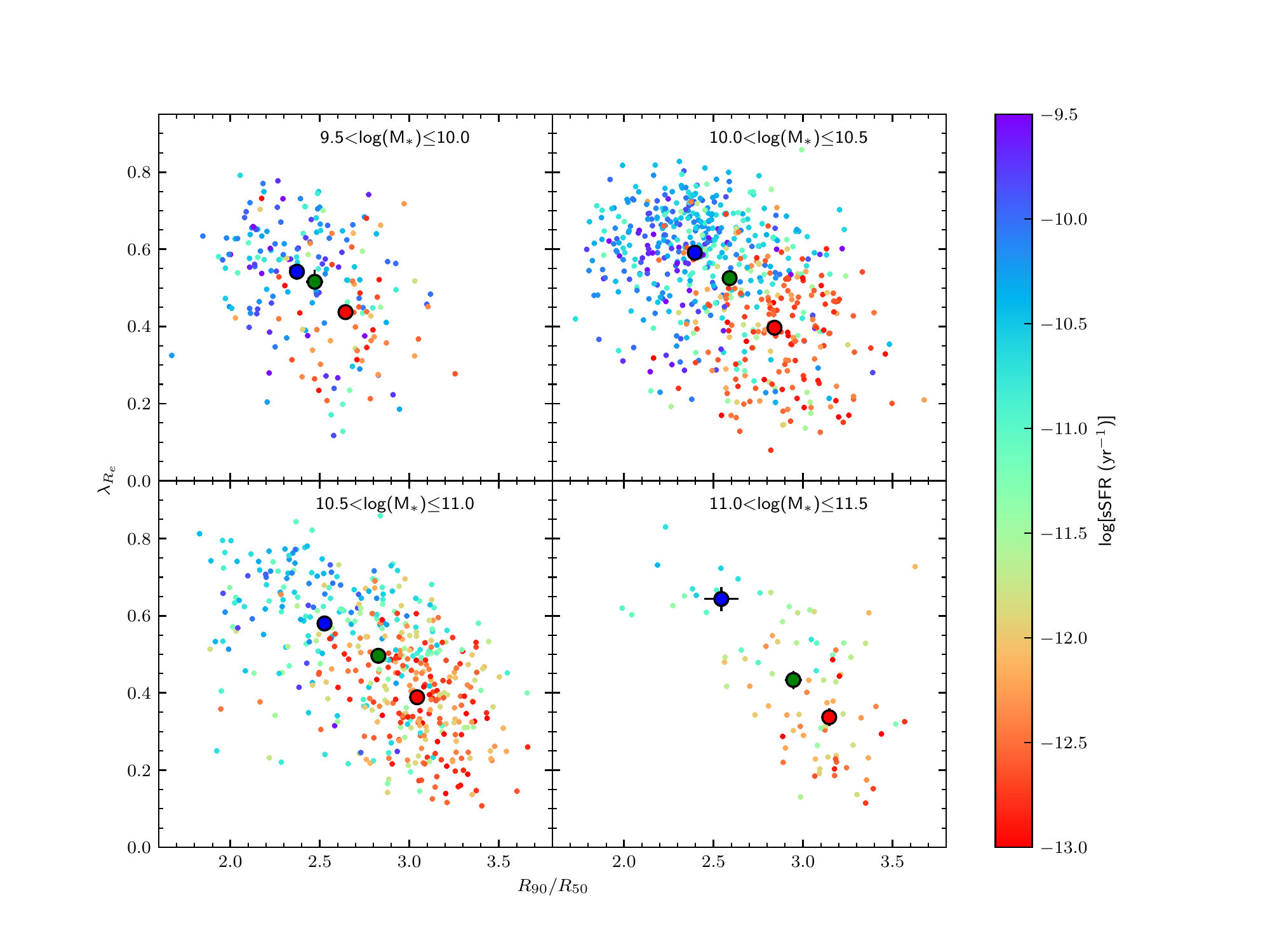}
    \caption{The distribution of our SAMI galaxies in the $\lambda_R$--concentration plane.  Galaxies are separated into 0.5 dex intervals of stellar mass and then colour--coded by specific star formation rate (sSFR).  The large points show the mean $\lambda_R$ and concentration values for the SF (blue), INT (green) and PAS (red) galaxy samples that are defined based on their location with respect to the star forming main sequence. The error bars on the large points show the error on the mean and are usually smaller than the points.}
    \label{fig:lam_conc_sami_sfr}
\end{figure*}

\subsection{Trends in $\lamre$ and concentration for star--forming and passive fast rotators}

Another consideration when looking at our comparison between eSp and S0 galaxies is whether optical morphology is the appropriate way to categorize such galaxies.  There are clearly significant structural and kinematic differences between eSp galaxies and S0 galaxies.  However, classifying these galaxies based on the presence of spiral arms or a bulge may only be an indirect approach to identifying the primary physical differences between these populations.  In almost all cases spiral arms will also signify star formation, as the enhanced star formation in the spiral arms leads to easier identification.  The connection between morphology and star formation is well known \citep[e.g.][]{2019MNRAS.483.1881D}.  However, there are cases of redder spirals that have weak or undetected star formation \citep[e.g.][]{2019ApJ...880..149P}.  A more direct approach might be to simply compare star-forming and passive fast-rotating galaxies to assess the degree of dynamical evolution as the star formation is quenched.

To estimate current star formation in our SAMI galaxies we use the $H\alpha$ emission line maps.  We remove regions that have non-star forming line ratios in the \oiii/H$\beta$ vs. \nii/H$\alpha$ ionization diagnostic diagram \citep{1981PASP...93....5B}, i.e. those spaxels that are significantly ($>1\sigma$) above the line defined by \citet{2003MNRAS.346.1055K}.  We then sum the star-forming H$\alpha$ flux and convert to star formation rate following a prescription similar to \citet{2018MNRAS.475.5194M}.  The H$\alpha$ flux is corrected for internal extinction using a Balmer decrement averaged across the galaxy and then converted to a star formation rate using the conversion of \citet{1994ApJ...435...22K}, but corrected to assume a \citet{2003PASP..115..763C} initial mass function.  This estimate of star formation may underestimate the total star formation rate in some cases, such as when the star-forming disc extends beyond the SAMI field of view, or when non-star forming spaxels rejected by ionization diagnostics still contain some star formation.  However,  our goal here is to broadly classify galaxies as star-forming or passive, and we find that our results are not sensitive to the details of the star formation estimate.

To classify our sample we use a galaxy's location with respect to the star-forming main sequence.  We use the local relation defined by \citet{2015ApJ...801L..29R} such that $\log[SFR_{\rm MS}/(M_* {\rm yr}^{-1})]=(0.76\pm0.01)\log(M_*/M_\odot)-(7.64\pm0.02)$ with a scatter of 0.3 dex.  We then define $\Delta(SFR) = \log(SFR_{\rm MS})-\log(SFR)$.  Our star forming population (labelled SF) is defined as galaxies that have $\Delta(SFR)>-0.6$ (twice the measured scatter).  We define an intermediate population (labelled `INT') with $-1.6<\Delta(SFR)<-0.6$, and a passive population (labelled `PAS') with $\Delta(SFR)<-1.6$.  Our conclusions are not sensitive to the specific thresholds that we set for these categories.  

The distribution of galaxies with $\lamre$ measurements in the specific star formation rate (sSFR) vs. mass plane can be seen in Fig.\ \ref{fig:sfr_ms}.  With the points in Fig. \ref{fig:sfr_ms} colour-coded by $\lamre$ a strong trend is visible, with galaxies at low sSFR typically having much lower $\lamre$.  We classify galaxies as slow rotators based on the 
boundary defined by \citet{2016ARA&A..54..597C}.  These slow rotators are shown with black edges in Fig.\ \ref{fig:sfr_ms} and are predominantly passive, although we note a small number of slow rotators that sit on the SFR main sequence.

The $\lamre$, concentration and ellipticity values for the SAMI sample divided by SFR are listed in Table \ref{tab:sami_results_sfr} and displayed in Fig.\ \ref{fig:lam_conc_sami_sfr}.  Here we do not include objects classified as slow rotators, under the assumption that they follow a different formation pathway.  The points in Fig.\  \ref{fig:lam_conc_sami_sfr} are colour coded by specific star formation rate (sSFR).  At stellar masses above $\log (M_*/M_\odot)=9.5$ there is a clear trend for galaxies with decreasing sSFR to have lower $\lamre$ and higher concentration.  This trend is as expected given the trends we also see based on visual morphology.

Fig. \ref{fig:lam_conc_sami_sfr} shows that the mean $\lamre$ and concentration for galaxies classed as INT (large green points, with some residual star formation, but below the main sequence) are found between the SF galaxies (large blue points) and the PAS galaxies (large red points).  The differences are quantified in the lower two sections of Table \ref{tab:sami_delta}.  Comparing INT and SF galaxies we find $\Delta\lamre$ is  between $-0.03$ and $-0.21$, while $\Delta C$ is $\simeq0.1-0.4$.  Averaged over all masses ($9.5<\log(M_*/M_\odot)<11.5$) the difference in $\lamre$ between SF and INT galaxies is relatively small, with $\Delta \lamre=-0.081\pm0.011$, somewhat smaller than the eSp to S0 difference discussed above.  The change in concentration between SF and INT galaxies, of $\Delta C=0.309\pm0.026$, is also smaller than the eSp to S0 transition.

The difference between SF and PAS galaxies is much larger than that between SF and INT galaxies.  The average (over $9.5<\log(M_*/M_\odot)<11.5$) difference is $\Delta \lamre = -0.187\pm0.009$, approximately 2.4 times larger than the difference between SF and INT galaxies.  The concentration difference from SF to PAS galaxies is $\Delta C = 0.506\pm0.019$, 1.7 times larger than the difference between SF and INT galaxies.

\subsection{Comparison of SAMI data to disc fading models}

The changes that we see in the SAMI population can be directly compared to our disc fading models (see Figs. \ref{fig:lam_ellip} and \ref{fig:lam_conc}).  The mean SAMI values for eSp and S0 galaxies in the mass intervals above  $\log(M_*/M_\odot)=9.5$ (black points in Figs. \ref{fig:lam_ellip} and \ref{fig:lam_conc}) lie along the simulated disc fading tracks (coloured points).  The differences seen in the SAMI data are comparable to the disc fading models when we consider the most extreme model parameters.  The evolution in the disc fading models is largest for $B/T=0.5$.  With 60 degree inclination this leads to $\lamre$ changing by $-0.11$ for 5\,Gyr of disc fading.   Only in the highest mass interval ($\log(M_*/M_\odot)>11$) is the observed difference between eSp and S0 galaxies ($-0.288\pm0.044$) too large for the most extreme disc fading models.  Note that increasing the length of time that we fade the disc does not help very much, as most of the difference comes in the first few Gyr.  

Our most extreme disc fading model ($B/T=0.5$) gives a change on concentration of 0.13 for 5\,Gyr of fading, compared to an observed change of $0.417\pm0.026$ for the median over all masses (a factor of 3.2 larger).  Only the lowest mass interval, with $\Delta C=0.175\pm0.088$, gives changes in concentration that are consistent with this most extreme disc fading model.  

If we more realistically average over the range of possible disc fading models, then the overall impact of disc fading will be reduced.  Assuming a flat distribution in $B/T$ from 0 to 1, as well as a flat distribution of inclination, results in $\Delta\lamre=-0.055$ and $\Delta C = 0.082$.  These values are not strongly sensitive to the range that we average over.  If we average the models from $B/T=0$ to 0.5 we find $\Delta\lamre=-0.057$ and $\Delta C = 0.085$.  Comparing to the measured values from SAMI (Table \ref{tab:sami_delta}) we find that the disc fading is able to contribute $\sim50$ percent of the difference seen between eSp and S0 galaxies in $\lamre$, but only $\simeq20$ percent of the difference in concentration.  Therefore, while disc fading can make  substantial contribution to the observed difference between eSp and S0 galaxies, it is not able to account for all of the difference.

We should also consider that the models we have constructed provide us with the most optimistic level of disc fading.  That is, we have instantaneous quenching and a purely dispersion supported bulge.  If we take the $B/T$ ratios that give us the largest change, we still fail to find sufficient disc fading.  Further, the true $B/T$ distribution of the population is broad \citep[e.g.][]{2017A&A...597A..97M}, particularly for early--type galaxies and early spirals.  If all spirals can be transformed into S0s, those with lower bulge fraction (particularly late spirals) will not be so severely influenced by disc fading.   Our model also assumes the bulge is older than the disk.  Various works show that this may not always be the case \citep[e.g.][]{2018MNRAS.481.5580F}.

When we instead consider galaxies classed by their star formation properties we find a similar picture.  Averaging over all mass (Table \ref{tab:sami_delta}) our disc fading models can only contribute 30 percent of the change in $\lamre$ and 18 percent of the change in concentration between galaxies close to the star-forming main sequence (our SF sample) and those that are fully passive (our PAS sample).  Disc fading alone cannot change main sequence galaxies to passive fast rotators.

The difference between galaxies in the main sequence (our SF sample) and those with some remaining star formation (our INT sample) is smaller than the difference with the PAS sample.  In this case disc fading can contribute 71 percent of the change in $\lamre$ and 30 percent of the change in concentration.

The smaller difference between SF and INT galaxies is consistent with the picture discussed by \citet{2019MNRAS.485.2656C}.  By comparing mass matched samples of central and satellite galaxies within the SAMI Galaxy Survey they suggest that galaxies undergo little dynamical change until they are fully passive.    

The impact of disc fading for kinematics will also be reduced if the bulges have some rotation.  Decomposition of bulges shows that they have a range of dynamical properties.  With a sample of spirals and S0s, \citet{2012ApJ...754...67F} finds a diverse range of bulge kinematics, where bulges with higher S\'ersic index have lower rotation.  These objects are a mix of true and pseudo--bulges, with the pseudo--bulges having higher rotational support (although the two populations overlap). \citet{2018MNRAS.474.1307M} shows that within a sample of S0 galaxies from the CALIFA Survey there is also a diversity of bulge rotational support.  These results imply that the influence of disc fading we demonstrate is an upper limit.

\begin{figure}
	\includegraphics[width=\columnwidth]{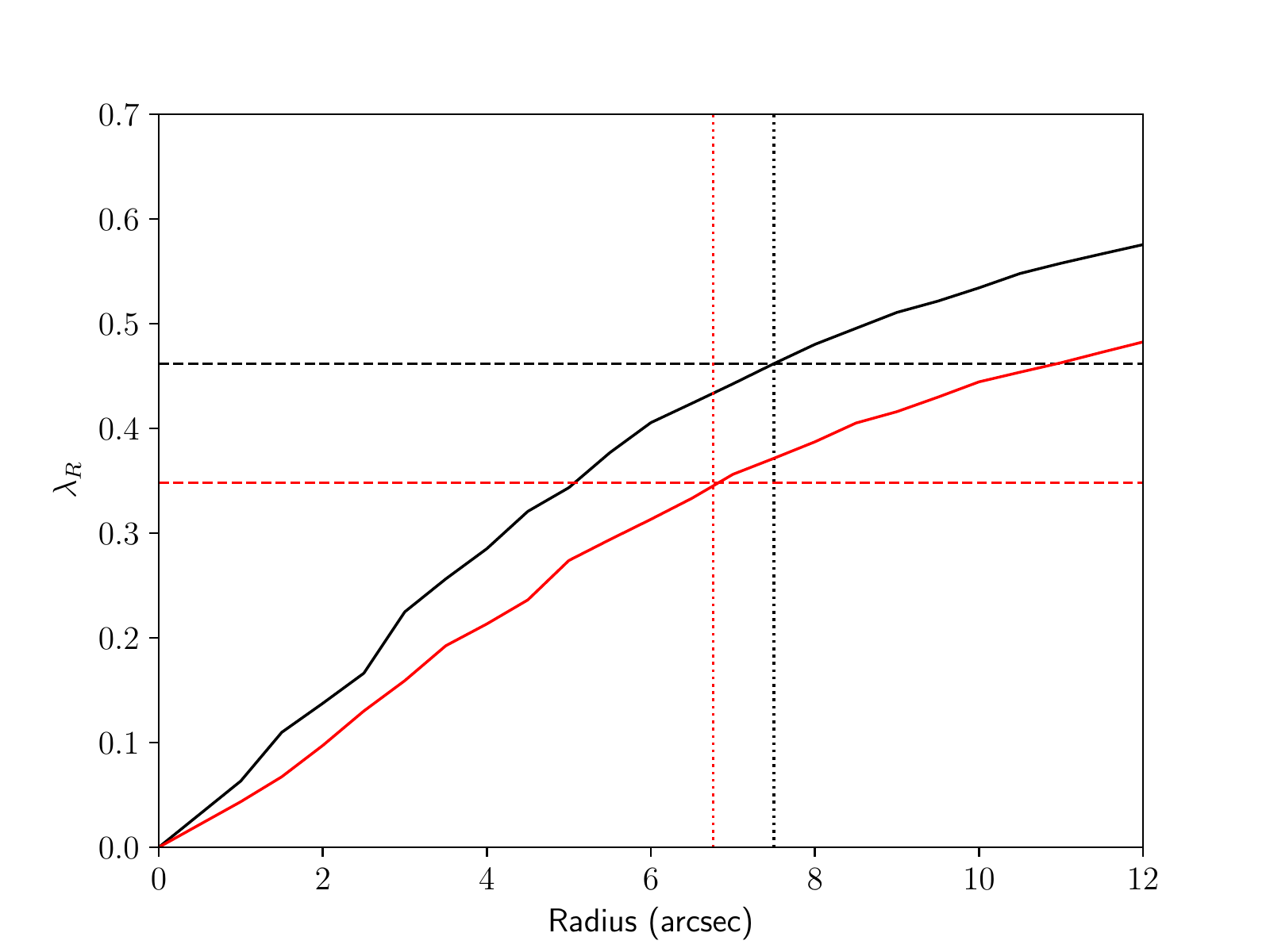}
    \caption{The $\lamr$ profiles (solid lines) for a model galaxy with $B/T=0.5$ and inclination of 60 degrees, 0\,Gyr (black) and 5\,Gyr (red) after quenching.  The vertical dotted lines indicate the value of $\re$ for each model, while the horizontal dashed lines show the values of $\lamre$.}
    \label{fig:lam_profiles}
\end{figure}

\begin{figure*}
	\includegraphics[width=14cm]{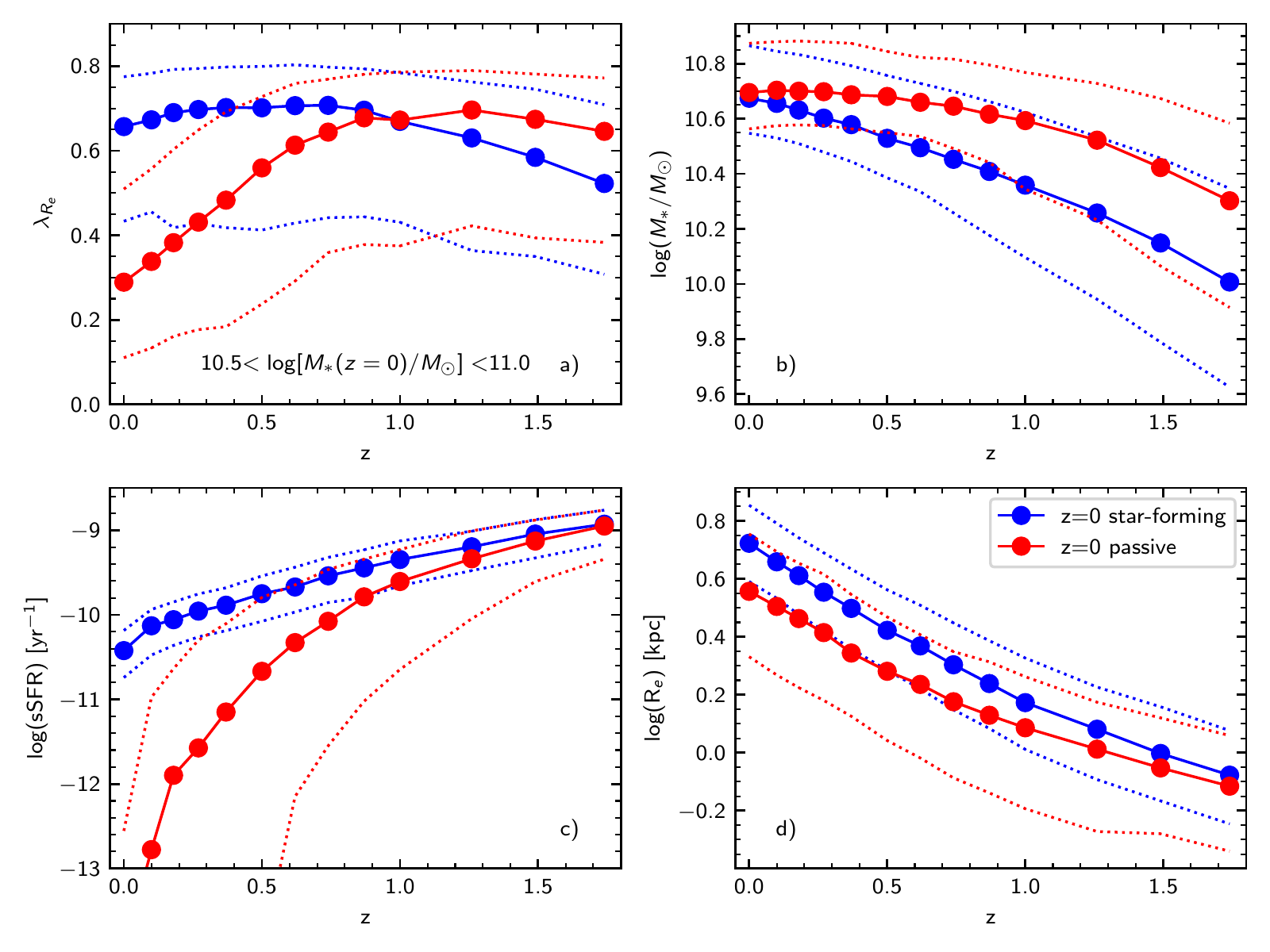}
    \caption{The median evolutionary tracks of galaxies selected from EAGLE to have $10.5<\log(M_*/M_\odot)\leq11$ at $z=0$.  We separately show galaxies selected to be on the star-forming main sequence (blue points) and those that are passive (red points) at $z=0$.  Dotted lines in each panel show the 68th percentile range of values.  We show the evolution for a) $\lamre$ (intrinsic edge-on); b) stellar mass; c) specific star formation rate; d) $r$-band half-light radius.}
    \label{fig:sim_evol}
\end{figure*}

\begin{figure*}
	\includegraphics[width=1.0\columnwidth]{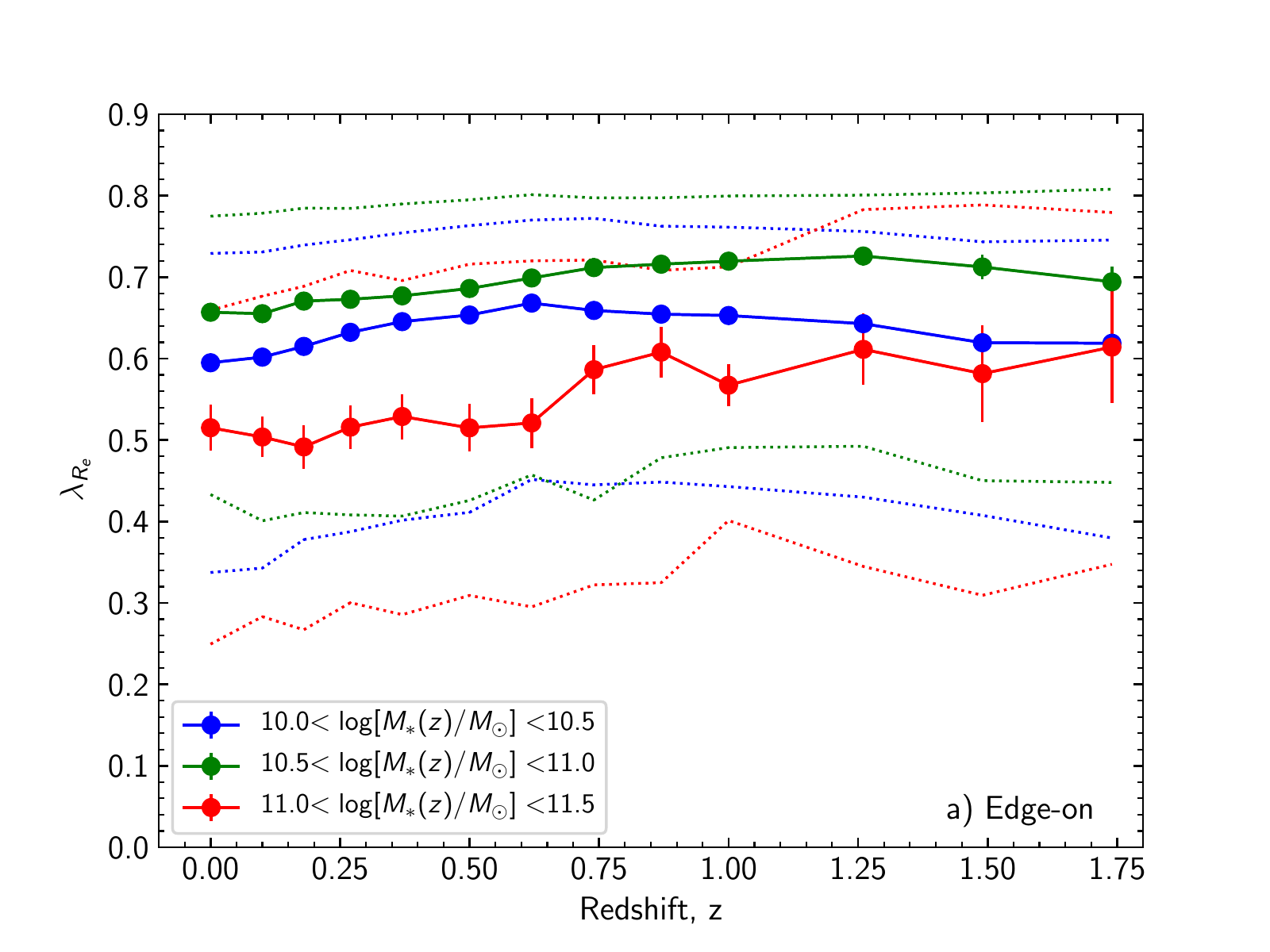}\includegraphics[width=1.0\columnwidth]{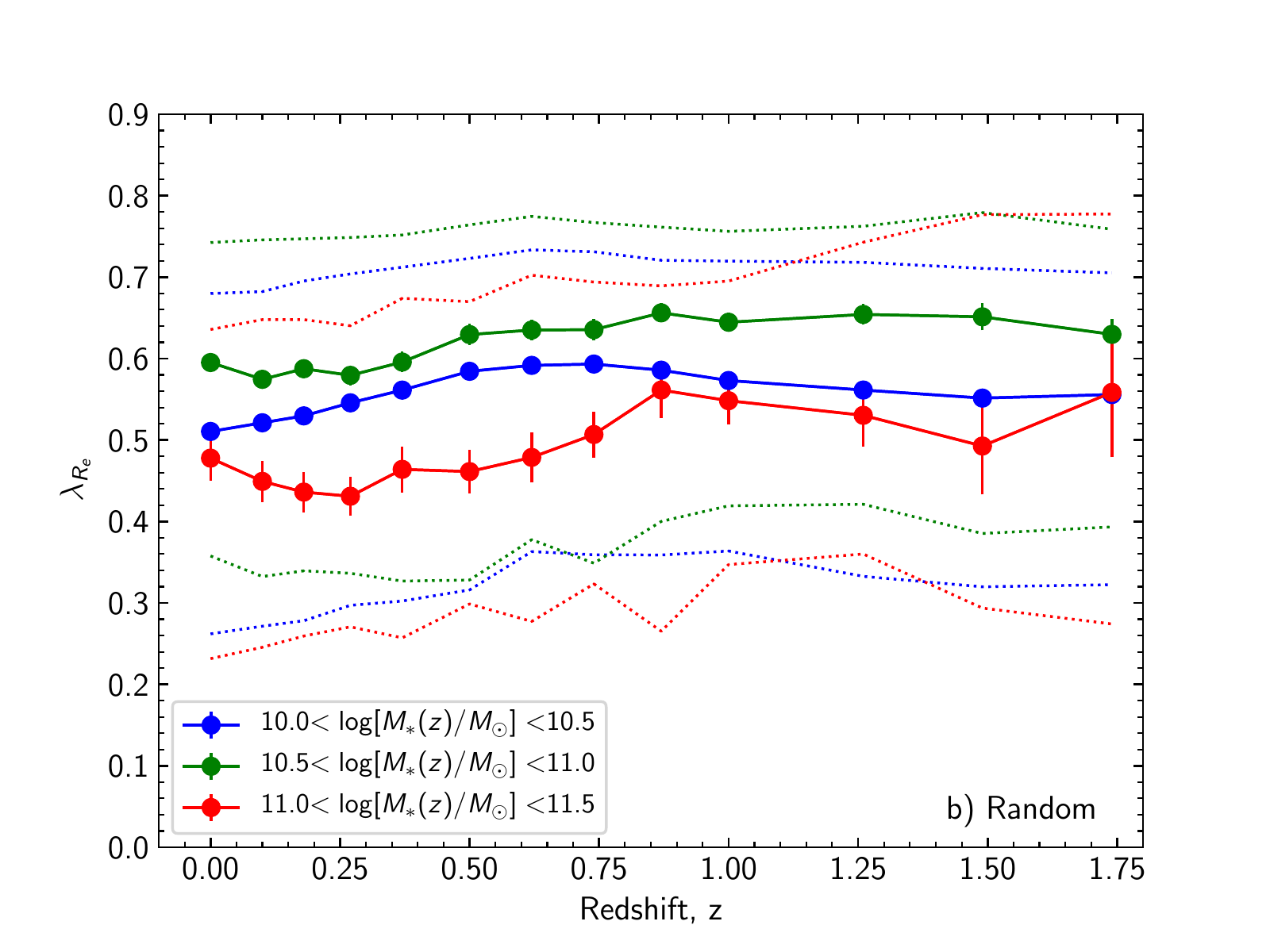}
    \caption{The simulated $\lamre$ for galaxies on the star-forming main sequence, based on measurements from EAGLE \citep{2015MNRAS.446..521S,2017MNRAS.464.3850L}.  Different colours corresponds to stellar mass intervals, where the stellar mass is measured at each different epoch.  Circles correspond to the median $\lamre$, the dotted lines show the 68th percentile range of values at the redshift of the measurement.  In a) we show results for the edge-on measurement of $\lamre$, while in b) we use random inclinations.  Errorbars show the error on the median that is typically smaller than the points for all but the highest mass.}
    \label{fig:sim_lam}
\end{figure*}

The contribution of disc fading to the observed kinematic change comes from two sources.  First is the larger photometric contribution to the bulge as the disc fades.  However, there is a second contribution from the evolution of the measured effective radius of the galaxy.  Given that the bulge has a smaller scale length than the disc, the overall $\re$ will get smaller as the disc fades.  As an example we show in Fig. \ref{fig:lam_profiles} the $\lamr$ profile for $B/T=0.5$ and inclination of 60 degrees.  The $\lamr$ profile is shown for a model with $t_q=0$ (black) and 5\,Gyr (red) and $\re$ is marked by the vertical dotted line.  In this case the change in $\lamr$ at a fixed radius is larger than the contribution due to changing $\re$.  This result is generally true for all models, but the exact contributions depend in detail on the model parameters.   

The above argues that while disc fading can be a significant contribution to the change in the measured kinematics and concentration, it is not sufficient to match the differences seen in our data.  Thus, other processes must also contribute significantly.  These processes could include dynamical heating via interactions with other galaxies or the potential of the parent halo.  However, another potential contribution could be from progenitor bias, which we will now discuss below.  

\subsection{The impact of progenitor bias}

An important caveat on the work above is that we have assumed that the progenitors of present day S0s (or PAS) galaxies look like present day eSp (or SF) galaxies.  As pointed out by \citet{2019MNRAS.485.2656C} and others, this need not be the case.  Thus, progenitor bias could also contribute to the observational differences that we see. In fact, observations of \ha\ emission suggest that in the past, galaxies had dynamically hotter discs \citep[e.g.][]{2015ApJ...799..209W}.  In the local Universe disks with younger stellar populations are thinner than those with older stars \citep[e.g.][]{2018NatAs...2..483V}.

\subsubsection{Progenitor bias from EAGLE simulations}

Galaxies that continue to accrete gas (and therefore continue to form stars) will tend to spin up with increasing cosmic time \citep[e.g.][]{2017MNRAS.464.3850L}.  To quantitatively assess the impact of progenitor bias we take measurements of $\lamre$ from the EAGLE simulations \citep{2015MNRAS.446..521S,2017MNRAS.464.3850L}. {\update We note that EAGLE size evolution \citep{2017MNRAS.465..722F} is consistent with the observed size evolution found by  \citet{2014ApJ...788...28V}.  As a result, this realistic size evolution will be implicitly included in the EAGLE $\lamre$ measurement, given that they are made within one effective radius in the $r$-band.}  We use the EAGLE reference model (Ref-L100N1504) and measure $\lamre$ as described by \citet{2018MNRAS.476.4327L} {\update and make measurements at 13 redshift intervals between $z=0$ and $1.8$.  Here and below we will only focus on measurements of $\lamre$ (not concentration) as we are primarily concerned with the evolution of kinematics.} 

{\update The galaxies in EAGLE are separated into passive and star forming, using a similar approach to the one we use with SAMI galaxies.  This is also similar to the method used on EAGLE galaxies by \citet{2019MNRAS.487.3740W}.  We first define potential star forming galaxies in EAGLE as those above $\log(sSFR/{\rm Gyr}^{-1})>-2+0.5z$ following \citet{2015MNRAS.450.4486F}.  Then, to define the star-forming main sequence for each redshift interval, we fit a linear relation to the median $\log(SFR)$ as a function of $\log(M_*)$.  EAGLE galaxies within 0.6 dex of the best fit main sequence relation are defined as star forming.  Those more than 1.6 dex below the main sequence are defined as passive.}

{\update Examples of the median EAGLE evolutionary tracks are shown in Fig.\ \ref{fig:sim_evol}.  Here we select EAGLE galaxies at $z=0$ that have $\log(M_*/M_\odot)=10.5-11$ and are either star forming (blue points/lines) or passive (red points/lines) at $z=0$.  We then identify their progenitors at higher redshift and calculate the median values of $\lamre$ (intrinsic edge-on value), stellar mass, specific star formation rate and $r$-band half-light radius.  From high redshift the $\lamre$ of the $z=0$ selected star-forming galaxies is increasing, but the evolution flattens below $z\sim1$.  In contrast, the passive galaxies decline in spin below $z=1$.  At $z>1$ the progenitors of $z=0$ passive galaxies have higher $\lamre$ than star-forming galaxies.  The reason for the offset in $\lamre$ at high $z$ can be explained by viewing the other panels in Fig.\ \ref{fig:sim_evol}.  Even though the progenitors of both the star-forming and passive galaxies have similar specific star formation rates above $z=1$ (see Fig.\ \ref{fig:sim_evol}c), they have different masses (Fig.\ \ref{fig:sim_evol}b), as the mass growth of passive galaxies is slower at $z<1$.  As a result, the progenitors of the $z=0$ star-forming and passive galaxies have different masses at high redshift, and we should not expect them to have the same $\lamre$ or other quantities (such as size, see Fig.\ \ref{fig:sim_evol}d), even if their specific star formation rates agree at an earlier epoch.
}

{\update The difference in $\lamre$ between star-forming and passive galaxies in EAGLE at $z=0$ is large, at $\sim0.25-0.35$ (depending on stellar mass).  This is similar to the difference we see in SAMI between passive and star-forming galaxies (see Table \ref{tab:sami_delta}). The large difference between star-forming and passive galaxies in the EAGLE simulations is also several times larger than the difference that can be attributed to disc fading.  The progenitors of the $z=0$ passive galaxies in EAGLE are also seen to decline in $\lamre$ by $\sim0.3$ between $z=1$ and $z=0$.  This decline is several times larger than can be attributed to disc fading, so the EAGLE simulations do not support simple disc fading as the cause of the low $\lamre$ in passive galaxies.}

{\update However, we note that the $\lamre$ distributions in different simulation data sets can be quite different \citep{2019MNRAS.484..869V}.  Also, as is highlighted by Fig.\ \ref{fig:sim_evol}, simply comparing the same mass galaxies at $z=0$ is not a sufficiently robust test to examine the importance of disc fading, as the mass growth histories of passive and star-forming galaxies are different.
}

{\update Using the EAGLE simulations we now examine the progenitors of today's passive galaxies by looking at the value of $\lamre$ for galaxies on the star-forming main sequence at different redshifts, but this time selected based on their stellar mass at redshift, $z$.}  We carry out this analysis in 3 stellar mass intervals, $10.0<\log(M_*(z)/M_\odot)\leq10.5$, $10.5<\log(M_*(z)/M_\odot)\leq11.0$ and $11.0<\log(M_*(z)/M_\odot)\leq11.5$.  In each case the masses correspond to the mass at the redshift where the properties are measured.  These mass intervals allow us to be sure that we have sufficient galaxies per bin (at high mass) and are not impacted by resolution effects (at low mass).  
The $\lamre$ values {\update on the main sequence} are shown in Fig. \ref{fig:sim_lam}.  We generate the measurements assuming that the galaxies are edge-on (Fig. \ref{fig:sim_lam}a) and randomly inclined to the line--of--sight (Fig. \ref{fig:sim_lam}b).  {\update For a given mass interval, $M_*(z)$, EAGLE galaxies on the main sequence have very similar median values of $\lamre$ at all redshifts we examine.  In fact, there is a small decline of up to $\sim0.1$ (for the highest mass interval) in $\lamre$ from high to low redshift.  We also note that for a given redshift, the value of $\lamre$ on the main sequence is a function of mass.  $\lamre$ increases from $10.0<\log(M_*(z)/M_\odot)\leq10.5$ (blue lines in Fig.\ \ref{fig:sim_lam}) to $10.5<\log(M_*(z)/M_\odot)\leq11.0$ (green lines).  However, as we increase mass to $\log(M_*(z)/M_\odot)>11.0$ $\lamre$ is lower again (red lines).  This is consistent with the increased importance of mergers in mass growth at the highest stellar masses.
}

\begin{figure}
	\includegraphics[width=1.1\columnwidth]{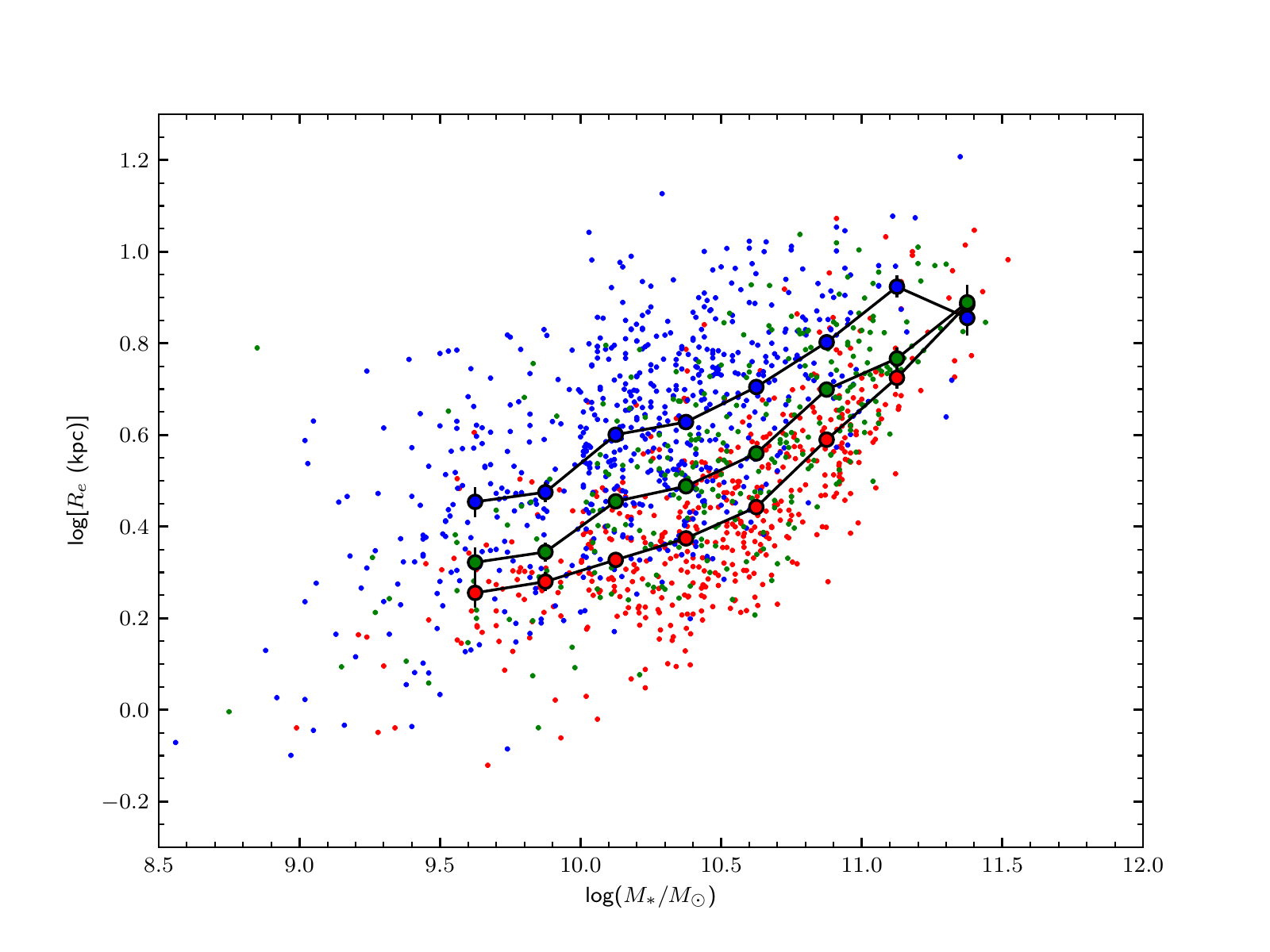}
    \caption{SAMI galaxies that we use in our analysis in the size (major axis $\re$ in SDSS $r$-band) vs.\ mass plane.  Points are colour coded based on their star formation classification as SF (blue), INT (green) or PAS (red).  The large connected points with errorbars denote the mean sizes in 0.25 dex stellar mass bins.  We do not include slow rotators in this analysis.}
    \label{fig:size_mass}
\end{figure}

\begin{figure}
	\includegraphics[width=1.1\columnwidth]{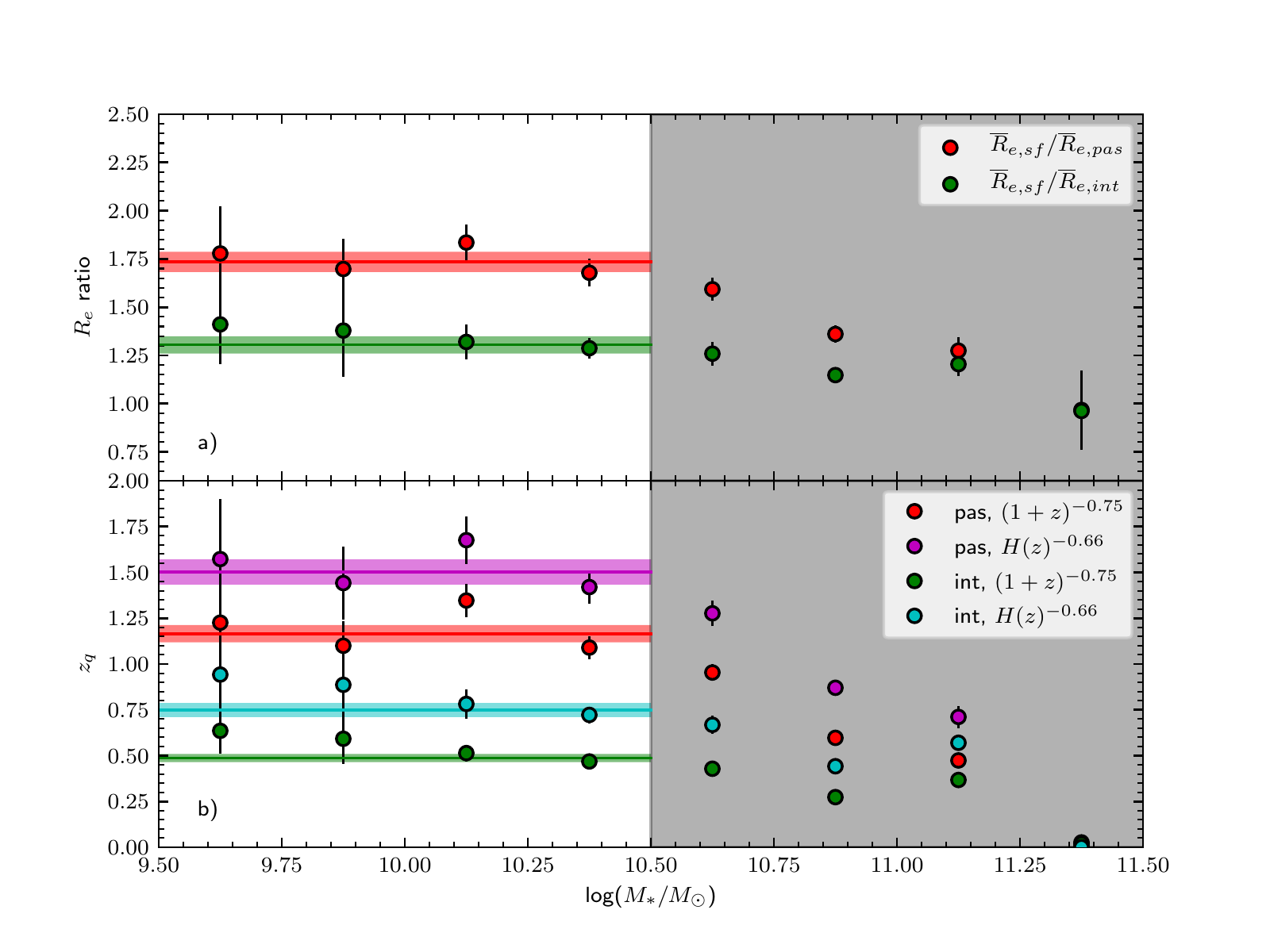}
    \caption{a) the ratio of mean size as a function of stellar mass for SAMI galaxies used in our analysis.  Red points denote $R_{\rm e, SF}/R_{\rm e, PAS}$ and green points denote $R_{\rm e, SF}/R_{\rm e, INT}$.  b) The inferred quenching redshift, $z_q$, for PAS and int galaxies based on their size ratios compared to SF galaxies.  We estimate $z_q$ using two different fitted forms for size evolution from \citet{2014ApJ...788...28V} where the evolution is either a function of $(1+z)$ or $H(z)$.  The grey shaded region indicates the mass range where mergers dominate mass growth, so we do not consider these points.  Horizontal lines are the average (over stellar masses $\log(M_*/M_\odot)<10.5$) $z_q$ for each sample and assumed evolution.}
    \label{fig:mass_size_evol}
\end{figure}

\subsubsection{Progenitor bias and size evolution}

The known \citep[e.g.][and references therein]{2014ApJ...788...28V} size evolution observed in the high redshift galaxy population also provides another source of progenitor bias.    To quantify this we compare the size--mass relations for SAMI galaxies used in our analysis, separated into the SF, INT and PAS populations (not including slow rotators; see fig.\ \ref{fig:size_mass}).  Here we are using the major axis $\re$ values estimated using MGE in the SDSS $r$-band.  There is a clear separation in size between SF and PAS galaxies that is largest at the low mass end.  This is not surprising given that the size--mass relation for early-type galaxies is known to be steeper than that of late types \citep[e.g.][]{2003MNRAS.343..978S}.  The INT galaxies sit between the SF and PAS populations.

Taking a step further we can use the ratio of mean size for SF and PAS or INT galaxies to quantify how much size {\update difference} there is between the populations.  These ratios are shown in Fig.\ \ref{fig:mass_size_evol}a.  At stellar masses below $\log(M_*/M_\odot)<10.5$ the size ratio is relatively constant, while at higher masses (grey shaded region) it declines.  Several authors \citep[e.g.][]{2014MNRAS.444.3986R} have pointed out that at masses greater than $\sim M^*$ galaxy build up is dominated by merging, while at lower masses in-situ star formation dominates.  Given that locally $\log(M^*/M_\odot)\simeq10.66$ \citep{2012MNRAS.421..621B}, we will only consider size information below $\log(M^*/M_\odot)=10.5$.  The average size ratio for $R_{\rm e, SF}/R_{\rm e, PAS}$ is $1.75\pm0.05$ (red horizontal line in Fig.\ \ref{fig:mass_size_evol}), while for $R_{\rm e, SF}/R_{\rm e, INT}$ it is $1.30\pm0.04$ (green horizontal line in Fig.\ \ref{fig:mass_size_evol}).  In comparison to these values, the change in size inferred by our disc fading models is only 10 percent at most (with $B/T=0.5$, see Fig.\ \ref{fig:re_tq}).  This strongly rules out the simple proposition that the progenitors of today's passive galaxies are the same as today's star forming galaxies.  The intermediate galaxies are also inconsistent with this proposition.  These size differences mean that progenitor bias must play an important role.

\subsubsection{\update{Using size evolution to infer quenching time}}\label{sec:size_quench}

Estimates of size evolution have previously been published by a number of authors including \citet{2014ApJ...788...28V}.  They find that for star-forming galaxies the zero-point evolution of the size-mass relation can be fit as either $\sim(1+z)^{-0.75}$ or $\sim H(z)^{-0.66}$, where $H(z)$ is the Hubble parameter at redshift $z$.  If we make the simplifying assumption that no further structural change occurs after a galaxy starts to leave the star--forming main sequence, then the size evolution model can be used to infer the redshift at which the galaxy quenched (which we will call $z_{\rm q}$).  For example, the value of $z_q$ for the PAS population assuming evolution parameterized by $(1+z)$ we find
\begin{equation}
z_{\rm q} = (1+z_{\rm obs})\left(\frac{\overline{R}_{\rm e,SF}}{\overline{R}_{\rm e, PAS}}\right)^{1/0.75} -1,
\label{eq:sizeevol}
\end{equation}
where $z_{\rm obs}$ is the redshift at which the SAMI galaxies are observed.  What we are effectively doing here is to use the fitted evolutionary trends from \citet{2014ApJ...788...28V} to find the redshift at which star--forming main sequence galaxies had the same size as the PAS {\update or} INT galaxies within the SAMI sample.  

In the above we are assuming that there is no further size evolution (other than due to disk fading) once disk galaxies have quenched.  However, observationally we know that size evolution of the passive population is stronger than for the star forming population.  This strong evolution of the passive population is partly driven by a real decline in the density of compact galaxies as we move to the present day \citep{2014ApJ...788...28V}, particular at high mass ($>5\times10^{10}$\,M$_\odot$).  At lower mass, it is plausible that most of the size evolution of the passive population is driven by the addition of larger star-forming disk galaxies as they quench.  As a result, it does not make sense to incorporate the measured size evolution of the passive population into our model.  We note that invariably the estimate of $z_{\rm q}$ based on size is only approximate.  We are not aiming to construct a detailed model that takes into account all potential elements of size evolution.

{\update
In summary the steps we take to estimate $z_{\rm q}$ from size evolution are as follows:
\begin{enumerate}
\item Find the size ratio between the observed SF and PAS (or INT) SAMI galaxy populations for each mass interval (see Fig.\ \ref{fig:mass_size_evol}a).
\item Assume there is no intrinsic dynamical evolution in a galaxy following its quenching and that size evolution due to disk fading is not significant.
\item Use fits to the observed size evolution \citep{2014ApJ...788...28V} to infer the redshift at which quenching takes place (see Eq. \ref{eq:sizeevol} and Fig.\ \ref{fig:mass_size_evol}b).
\end{enumerate}
}

The results of our $z_{\rm q}$ calculation can be seen in Fig.\ \ref{fig:mass_size_evol}b and here we again only consider galaxies at $\log(M_*/M_\odot)<10.5$.  For PAS galaxies the inferred mean $z_{\rm q}$ is $1.18\pm0.05$ and $1.53\pm0.07$ (for the $(1+z)$ and $H(z)$ evolution parameterizations respectively).  The difference between the $(1+z)$ and $H(z)$ evolution estimates of $z_{\rm q}$ is because the models diverge somewhat at $z=0$ [see Fig.\ 6 of \citet{2014ApJ...788...28V}].  Other estimates of size evolution, including a constraint at $z\simeq0$ from  SDSS find a solution that is closer to the $H(z)$ model \citep{2006ApJ...650...18T}, however for completeness we consider both parameterizations.  For INT galaxies the $z_{\rm q}$ values are $0.48\pm0.02$ and $0.74\pm0.04$.  The mean values below $\log(M_*/M_\odot)=10.5$ are indicated in Fig.\ \ref{fig:mass_size_evol}b by the horizontal lines.  It is clear from these values and Fig.\ \ref{fig:mass_size_evol}b that uncertainties in the evolutionary model contribute significantly to the calculated $z_{\rm q}$.  Residual size evolution (e.g.\ due to disc fading, see Fig.\ \ref{fig:re_tq}) could also contribute to uncertainty on $z_{\rm q}$, but as we discuss above, this is small compared to the overall evolution in size seen in the galaxy population.

\subsubsection{\update{Combining disc fading and progenitor bias}}

\citet{2016ApJ...818..180C} demonstrated from photometric measurements that the differences in $B/T$ between quenched and star-forming satellite galaxies can be largely attributed to disc-fading.  However, Carollo et al.\ also show that the size difference between quenched and star forming galaxies is too large to be caused by disc fading.  This is consistent with our measurements, that also consider dynamics.  

The morphological mix of galaxies in groups is known to evolve strongly, with an increase in the fraction of S0s by approximately a factor of 2 since $z\simeq0.5$ \citep{2010ApJ...711..192J}.  {\update If we assume there is little mass growth in passive disks once they are quenched (see Fig.\ \ref{fig:sim_evol}b and c), then the value of $\lamre(z)$ for main sequence galaxies in Fig.\ \ref{fig:sim_lam} provides us an estimate of their progenitors' spin at the point that they quench.  If we conservatively say that most S0s have transformed from star-forming discs since $z\sim1$, then the EAGLE simulation results suggest that their progenitors had $\lamre$ that is slightly higher than current star-forming discs of the same mass.  For example, the difference $\lamre(z=0)-\lamre(z=1)=-0.06$, -0.05 and -0.07 for EAGLE main sequence galaxies in mass intervals $\log(M_*/M_\odot)=10-10.5$, $10.5-11.0$ and $11.0-11.5$ respectively.  These values are similar to the change expected from 5\,Gyr of our disc fading models ($\Delta\lamre=-0.055$, averaged over all B/T and inclination).  As a result, disk fading is only sufficient to evolve galaxies from their typical $\lamre$ on the main sequence at $z=1$ to the typical value of $\lamre$ on the main sequence at $z=0$.  This is not a sufficient change to evolve galaxies on the main sequence to $z=1$ to the observed $\lamre$ of passive galaxies at $z=0$.}
 
We can also use our estimates of $z_q$ from size evolution to obtain another prediction of the amount of progenitor bias in $\lamre$.  For PAS galaxies, $z_q=1.13-1.47$ (depending on the evolutionary model used). {\update  At the high end of this redshift range the $\lamre$ of galaxies on the main sequence (Fig.\ \ref{fig:sim_lam}) starts to decline towards higher redshift, however this decline is small.  As a result, for all three mass intervals shown in Fig.\ \ref{fig:sim_lam}, $\lamre$ in the main sequence is higher at $z_q=1.13-1.47$ than it is at $z=0$.}

{\update We consider our size evolution estimates of quenching time to be reasonable at $\log(M_*/M_\odot)=10-10.5$ (Section \ref{sec:size_quench} and Fig.\ref{fig:mass_size_evol}).  In this mass range the $\lamre$ on the main sequence is between 0.06 ($z_q=1.13$) and 0.04 ($z_q=1.47$) higher than at $z=0$.  The observed difference between $z=0 $ PAS and SF galaxies in this mass range is $-0.194\pm0.012$.  If we add the progenitor bias to this, then the range for the total change in spin is between $-0.234$ and $-0.254$, where the size of the allowable range is dominated by the uncertainty on the estimated quenching time, not the uncertainty in observed $\lamre$ values.   In contrast, the change allowable due to disk fading (averaging over all $B/T$) is $-0.057$.  Disk fading appears to only contribute a small fraction of the required change, meaning that intrinsic dynamical evolution (i.e.\ changes in the orbital distribution of stars) must be an important contributing factor.  It is also worth noting that our disk fading models are likely to be optimistic in the amount of apparent kinematic change they cause. We assume a purely dispersion supported bulge (no rotation) and a 10\,Gyr old bulge stellar population.  A bulge with some rotation, or younger stars will reduce the impact of disk fading.}

{\update The galaxies from the EAGLE simulation on the SF main sequence at $\log(M_*/M_\odot)>11$ have the largest increase in $\lamre$ as redshift increases.  These galaxies also show the largest observational difference in $\lamre$ between SF and PAS (or eSp and S0s).  In this case it is even more clear that intrinsic dynamical evolution must play the main role in transforming galaxies.}  At high masses mergers are more important than in--situ star formation for mass growth \citep[e.g.][]{2014MNRAS.444.3986R}.  Therefore, we expect that many of the passive fast rotators at $\log(M_*/M_\odot)>11$ may be built up from mergers.  These mergers will need to have impact parameters and total angular momentum such that after the merger, they still have significant rotation.  \citet{2018MNRAS.476.4327L} used EAGLE simulations to show that galaxy mergers consistently decrease $\lamre$ unless they are very gas rich.  Minor mergers with gas fraction $<0.1$, can reduce $\lamre$ by 20 to 40 percent on average, while major mergers of the same gas fractions reduce $\lamre$ by 50 percent [see Fig.\ 14 of \citet{2018MNRAS.476.4327L}].  Hence, the decrease in $\lamre$, even of main sequence galaxies, is consistent with mergers affecting them systematically at $z<1$.

To confirm the contribution of progenitor bias we will need to make stellar kinematic observations at higher redshift.  Some work in this area has already been done using the LEGA-C survey \citep{2018ApJ...858...60B}.  {\update LEGA-C finds that $V/\sigma$ for passive galaxies is reduced from $z=0.8$ to $z=0$, suggesting some spin down of massive passive galaxies.  However, the exact amount of spin down is still uncertain [e.g.\ see discussion in appendix of \citet{2018ApJ...858...60B}].  These measurements are also seeing-convolved estimates of $V/\sigma$ evolution and seeing corrections are dependent on both $V/\sigma$ and size \citep{2020MNRAS.tmp.1978H}, so the intrinsic evolution is harder to discern.}

\section{Conclusions}\label{sec:conc}

We use dynamically self consistent models to estimate the signature of disc fading on the observed kinematics and structural properties of galaxies.    In particular, we assess how the changing contribution from bulge and disc can influence galaxy properties, despite no change in the mass fraction in each component.  Specifically we conclude that:

\begin{enumerate}

\item In galaxies with an old bulge and a star forming disc, quenching the disc leads to a reduction in measured spin, $\lamre$ and an increase in measured concentration.   These trends are due to the reduction in the light weighted contribution of the disc.  Unsurprisingly this is most significant for systems where the mass of the bulge and disc are relatively equal.  For $B/T=0.5$ we find that 5\,Gyr of quenching leads to a reduction of $-0.12$ in $\lamre$.  At the same time, the measured $r$-band concentration increases by 0.13.  Averaged over all $B/T$ we find 5\,Gyr of disc fading leads to change in $\lamre$ of $-0.055$ and a change in concentration of 0.082.

\item We measure the difference in $\lamre$ and concentration between early spirals (classified eSp) and S0s from the SAMI Galaxy Survey.  We find S0s have an average $\lamre$ that is  smaller than eSp galaxies, with the mean $\Delta\lamre=-0.132\pm0.013$.  The mean difference in concentration is $\Delta C = 0.417\pm0.026$.  This difference is in the same qualitative sense as our disc fading models but somewhat larger in amplitude, only becoming comparable when using the most extreme models ($B/T=0.5$).   

\item When we separate SAMI galaxies by their star formation rate relative to the main sequence (instead of morphology) we find that the difference between regular star forming galaxies (on the main sequence) and passive galaxies ($>1.6$\,dex below the main sequence) is too great to be due to disc fading alone.

\item The difference in spin and concentration between main sequence galaxies and those with weak star formation ($0.6-1.6$\, dex below the main sequence) is less than for passive galaxies.

\item Size evolution plays an important role in progenitor bias, but as a result can be used as a tool to estimate the time at which galaxies quenched and left the main sequence (under a number of assumptions).

\item We use the EAGLE simulations to estimate the amount of progenitor bias that can contribute to $\lamre$ differences.  {\update For a fixed mass range at redshift $z$, the spin of main sequence galaxies increases slightly from $z=0$ until at least $z=1$ (dependent on mass).  This progenitor bias does not help to bring disk fading models into agreement with the data, as it goes in the opposite sense to that required.  We conclude that disk fading is not sufficient to explain the $\lamre$ difference between star forming and passive (or eSp and S0) galaxies at $z=0$.  Instead, intrinsic evolution of the stellar dynamics of the galaxies must dominate.}

\end{enumerate}

The progenitors of today's S0s are {\update in most cases likely to be spiral galaxies at an earlier epoch \citep[e.g.][]{1997ApJ...490..577D,2010ApJ...711..192J}}.  Spatially resolved spectroscopic surveys capable of measuring the stellar kinematics in discs should be able to quantify the importance of evolving stellar disc dispersion.  Sampling a redshift range $z=0.25-0.50$ would be sufficient to study 3--5\,Gyr of evolution.  Such observations are now becoming possible with medium--deep surveys covering large areas using instruments such as the Multi Unit Spectroscopic Explorer \cite[MUSE;][]{2010SPIE.7735E..08B}.  One such project is the Middle-Ages Galaxy Properties with Integral Field Spectroscopy {\update \citep[MAGPI;][]{2020arXiv201113567F}} survey currently proceeding on the Very Large Telescope with MUSE.

Future work should also examine whether the kinematic and structural properties of S0 galaxies are dependent on environment.  If  dynamical effects contribute to the formation process of S0s, differences in interactions as a function of environment could lead to measurable environmental trends. 

\section*{Acknowledgements}

The SAMI Galaxy Survey is based on observations made at the Anglo-Australian Telescope. The Sydney-AAO Multi-object Integral field spectrograph (SAMI) was developed jointly by the University of Sydney and the Australian Astronomical Observatory. The SAMI input catalogue is based on data taken from the Sloan Digital Sky Survey, the GAMA Survey and the VST ATLAS Survey. The SAMI Galaxy Survey is supported by the Australian Research Council Centre of Excellence for All Sky Astrophysics in 3 Dimensions (ASTRO 3D), through project number CE170100013, the Australian Research Council Centre of Excellence for All-sky Astrophysics (CAASTRO), through project number CE110001020, and other participating institutions. The SAMI Galaxy Survey website is http://sami-survey.org/ .

JvdS acknowledges support of an Australian Research Council Discovery Early Career Research Award (project number DE200100461) funded by the Australian Government and funding from Bland-Hawthorn's ARC Laureate Fellowship (FL140100278). JBH is supported by an ARC Laureate Fellowship and an ARC Federation Fellowship that funded the SAMI prototype.  JJB acknowledges support of an Australian Research Council Future Fellowship (FT180100231).  LC is the recipient of an Australian Research Council Future Fellowship (FT180100066) funded by the Australian Government.
M.S.O. acknowledges the funding support from the Australian Research Council through a Future Fellowship (FT140100255).  NS acknowledges support of an Australian Research Council Discovery Early Career Research Award (project number DE190100375) funded by the Australian Government and a University of Sydney Postdoctoral Research Fellowship.  KH is supported by the SIRF and UPA awarded by the University of Western Australia Scholarships Committee.

\section*{Data Availability}

The SAMI data used in this paper, including kinematic measurements, are included in SAMI Data Release 3 \citep{2021MNRAS.tmp..291C} available via Australian Astronomical Optics' Data Central, https://datacentral.org.au/.  Simulation results for all models are available online as part of this publication.




\bibliographystyle{mnras}
\bibliography{bibliographies} 








\bsp	
\label{lastpage}
\end{document}